\documentclass[iop,apj]{emulateapj}

\usepackage{amsmath}
\usepackage{color}

\newcommand{\cjaa}{Chin.~J.~of Astro.~and Astrophys.}

\shorttitle{galactic synchrotron emission and the fir-radio correlation at high redshift}
\shortauthors{Schober et al.}

\begin{document}

\author{J.~Schober}
\affil{Nordita, KTH Royal Institute of Technology and Stockholm University,\\ Roslagstullsbacken 23, 10691 Stockholm, Sweden}
\email{jschober@nordita.org}

\author{D.~R.~G.~Schleicher}
\affil{Departamento de Astronom\'ia, Facultad Ciencias F\'isicas y Matem\'aticas, Universidad de Concepci\'on,\\ Av. Esteban Iturra s/n Barrio Universitario, Casilla 160-C, Concepci\'on, Chile}

\author{R.~S.~Klessen}
\affil{Universit\"at Heidelberg, Zentrum f\"ur Astronomie, Institut f\"ur Theoretische Astrophysik,\\ Albert-Ueberle-Strasse~2, D-69120 Heidelberg, Germany}
\affil{Universit\"at Heidelberg, Interdisziplin\"ares Zentrum f\"ur Wissenschaftliches Rechnen,\\ Im Neuenheimer Feld~205, D-69120 Heidelberg, Germany}

\title{Galactic synchrotron emission and the FIR-radio correlation \\at high redshift}

\begin{abstract}
Galactic magnetic fields in the local Universe are strong and omnipresent. There is mounting evidence that galaxies were magnetized already in the early Universe. Theoretical scenarios including the turbulent small-scale dynamo predict magnetic energy densities comparable to the one of turbulence. Based on the assumption of this energy equipartition, we determine the galactic synchrotron flux as a function of redshift $z$. The conditions in the early Universe are different from the present day, in particular the galaxies have more intense star formation. To cover a large range of conditions we consider models based on two different systems: one model galaxy comparable to the Milky Way and one typical high-$z$ starburst galaxy. We include a model of the steady state cosmic ray spectrum and find that synchrotron emission can be detected up to cosmological redshifts with current and future radio telescopes. Turbulent dynamo theory is in agreement with the origin of the observed correlation between the far-infrared (FIR) luminosity $L_\mathrm{FIR}$ and the radio luminosity $L_\mathrm{radio}$. Our model reproduces this correlation well at $z=0$. We extrapolate the FIR-radio correlation to higher redshift and predict a time evolution with a significant deviation from its present-day appearance already at $z\approx2$ for a gas density that increases strongly with $z$. In particular, we predict a decrease of the radio luminosity with redshift which is caused by the increase of cosmic ray energy losses at high $z$. The result is an increase of the ratio between $L_\mathrm{FIR}$ and $L_\mathrm{radio}$. Simultaneously, we predict that the slope of the FIR-radio correlation becomes shallower with redshift. This behavior of the correlation could be observed in the near future with ultra-deep radio surveys.
\end{abstract}

\maketitle

\section{Introduction}
\label{sec_Introduction}

In order to understand the formation and evolution of galaxies, one needs to know the evolution of their main energy components. Observations show clearly that magnetic fields are present on all astrophysical length scales of present-day galaxies and are approximately in equipartition with the remaining energy components \citep{Beck2011}. However, they are often left out in galaxy modeling as they increase the complexity significantly. As star formation and large-scale galactic dynamics are influenced crucially by magnetic fields, it is necessary to know how magnetic fields have formed and evolved in time for solving the details of galaxy evolution. \\
Various scenarios for the origin of galactic magnetic fields have been proposed. The first seed fields could have been created in the very early Universe - during inflation 
\citep{TurnerWidrow1988,Ratra1992} or within phase transitions \citep{QuashnockEtAl1989,SiglEtAl1997}. However, due to cosmic expansion these fields have become extremely weak at the onset of galaxy formation. Stronger fields are expected from plasma mechanisms like the Biermann battery ($\approx 10^{-18}$ G, \citet{XuEtAl2008}) or from aperiodic plasma modes ($\approx 10^{-10}$ G, \citet{Schlickeiser2012}). At present day, galaxies like the Milky Way have field strengths of approximately $10^{-5}$ G \citep{Beck2001} which means that seed fields must have been amplified over many orders of magnitude during the galactic evolution. One of the best candidates for amplification is the turbulent small-scale dynamo. This process converts kinetic energy from turbulence into magnetic energy by randomly stretching, twisting, and folding field lines. As the dynamo operates, at least initially, on small spatial scales it is very fast \citep{BrandenburgSubramanian2005,FederrathEtAl2011b,SchoberEtAl2012.1,SchoberEtAl2012.3,FederrathEtAl2014b}. In the context of galaxy formation the turbulent dynamo has been studied with semi-analytical models \citep{SchoberEtAl2013} and also in numerical simulations \citep{LatifEtAl2013,PakmorMarinacciSpringel2014}. These studies show that equipartition field strengths can be reached on timescales of the order of $10^{6}$ yr \citep{SchoberEtAl2013}. We note, that the ordering of field lines as observed in galactic disks should take place on longer timescales which are typical for large scale motions. For instance the rotation period of the Milky Way is $2.4\times10^{8}$ yr. \\
Based on the recent studies \citep{SchoberEtAl2013,LatifEtAl2013,PakmorMarinacciSpringel2014} showing efficient dynamo amplification in young galaxies, we expect strong galactic fields also in the early Universe. To test this theoretical scenario we can study non-thermal radio emission from cosmic rays. While these highly energetic particles consist mostly of protons, the electrons are responsible for synchrotron emission due to their smaller mass. The origin of this emission is the spiral motion of the electrons around the magnetic field lines providing a probe of its strength and structure across a galaxy. This commonly used method is limited by the availability of information about the cosmic ray electron spectrum \citep{BeckKrause2005}. Theoretical models for the steady-state distribution of cosmic rays that are related to other galaxy properties like the star formation activity are needed to predict the synchrotron emission. In fact, with these models it is possible to also estimate radio fluxes from galaxies at different redshifts. \\
One hint towards a strong coupling between the different energy components in a galaxy is the far-infrared (FIR)-radio correlation. \citet{YunEtAl2001} observe this correlation of the FIR and the radio luminosity over more than five orders of magnitude. This tight correlation can be interpreted as a strong coupling between star formation and magnetic fields via cosmic rays \citep{SchleicherBeck2013}. The star formation rate is traced by the FIR luminosity that results from thermal emission of dust heated by UV radiation from young stars. For instance, \citet{Kennicutt1998} reports a linear scaling between the bolometric FIR luminosity, e.g.~integrated between 8 $\mu$m and 1000 $\mu$m, and the star formation rate. A higher star formation rate is also accompanied by the occurrence of more turbulence \citep{MacLowKlessen2004}. This leads to stronger magnetic fields produced by small-scale dynamo amplification. In addition, with the formation of more stars more supernova explosions occur which produce more cosmic rays in their shock fronts. We thus expect the non-thermal radio emission to increase with increasing star formation. The FIR-radio correlation is well-tested in the local Universe. While until today no clear evidence for an evolution of this correlation has been observed up to moderate redshifts \citep{MurphyEtAl2009a}, a deviation from its present-day shape is nevertheless discussed in literature \citep{Murphy2009,LackiEtAl2010b}. \\
The outline of this paper is as follows: We present our galaxy model in section \ref{sec_Model} and discuss our assumptions for the geometry, magnetic fields, and cosmic ray injection. As we aim to study a large range of star formation rates we base our model on two fiducial galaxies: the Milky Way and SMM J2135-0102 - a starburst galaxy at $z=2.3$. From our model we derive the steady-state spectrum of cosmic ray electrons. In section \ref{sec_Synchrotron} we employ our galaxy model to calculate the non-thermal radio luminosity emitted from the cosmic rays interacting with the galactic magnetic field. We compute the radio fluxes as a function of redshift and compare it to the sensitivities of current radio telescopes. In section \ref{sec_FIR-radio} we study the FIR-radio correlation resulting from our model. First, we compare our prediction for $z=0$ to local observations. Second, we calculate the correlation for higher redshifts and study its evolution in time. We summarize our findings in section \ref{sec_Conclusions}.


\section{Description of our model}
\label{sec_Model}

\subsection{General properties}
\label{subsec_GeneralProperties}

In this study we aim to predict the synchrotron emission of galaxies with various properties at different redshifts. In oder to cover a large range of star formation rates $\dot{M}_\star$ we employ two types of galaxies as our fiducial models: the Milky Way for low and moderate $\dot{M}_\star$ and SMM J2135-0102 for high $\dot{M}_\star$. The latter is a starburst galaxy at $z=2.3$. Due to the gravitational lensing effect it could be studied in detail \citep{IvisonEtAl2010}. We estimate unknown properties of SMM J2135-0102 using a local starburst, M 82. \\
In the following we refer to our two parameter regimes as ``normal disk'' and ``high-$z$ starburst''. We define the transition from the low to the high $\dot{M}_\star$ regime at $10~M_\odot\mathrm{yr}^{-1}$. The basic properties of our fiducial galaxies are summarized in table \ref{Table_Props}. In this section we discuss the choices of these values. Moreover, we construct a general model for a galaxy as a function of $\dot{M}_\star$ and the redshift $z$.

\begin{figure}
  \includegraphics[width=0.5\textwidth]{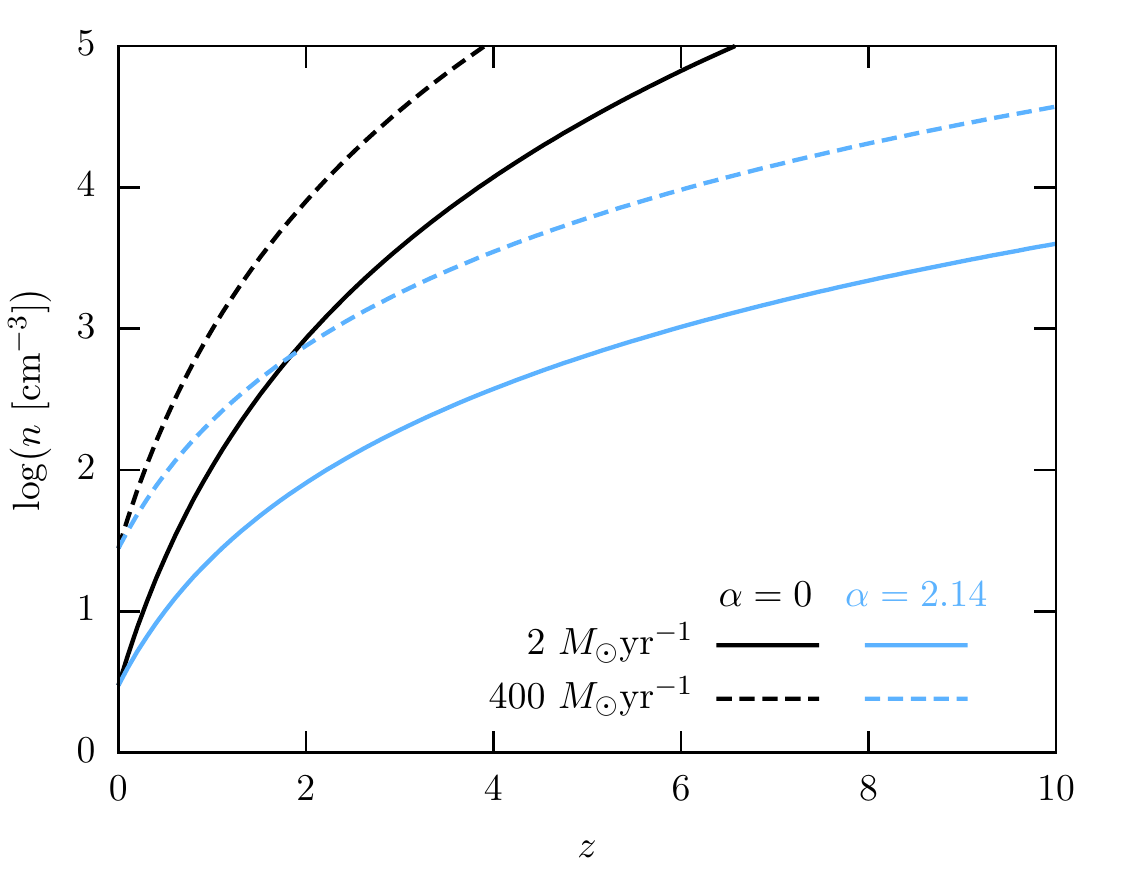}
  \caption{The gas density $n$ as a function of redshift $z$ for the Milky Way based model with a star formation rate of $2~\mathrm{M}_\odot\mathrm{yr}^{-1}$ (solid lines) and the high-$z$ starburst with $400~\mathrm{M}_\odot\mathrm{yr}^{-1}$ (dashed lines). We present a case of strong redshift evolution ($\alpha=0$) in black lines and one of weak redshift evolution ($\alpha=2.14$) in blue lines (see equation \ref{eq_n}).}
\label{plot_n_z}
\end{figure}

\paragraph{Volume and density of the neutral gas}
For our simple galaxy model we assume that the galaxies are disks with a uniform density. The fiducial values of the radii $R_0$ and the scale heights $H_0$ of this disks are given in table \ref{Table_Props}. We use typical values for the stellar disk of the Milky Way \citep{KennicuttEvans2012,RixBovy2013} and observations from \citet{IvisonEtAl2010} for the high-$z$ case. The typical scale height of a galaxy at high redshifts is not known. We assume the high-$z$ disks to be thicker than at present day where in our Milky Way model $R_0/H_0 = 30$. For SMM J2135-0102 we choose a lower ratio of $R_0/H_0 = 20$ leading to  $H_0 = 50$ pc at $z=2.3$.\\
With increasing redshift the disks are typically smaller. Under the assumption of a constant galactic mass semi-analytical calculations predict a scaling of the galactic radius proportional to $(z+1)^{-1}$ \citep{MoMaoWhite1998}. Observations of high redshift galaxies seem to agree with this scaling. In fact, by analyzing Hubble data up to redshifts of 8, \citet{OeschEtAl2010} find that the half-light radius evolves with $(1+z)^{-1.12 \pm 0.17}$. Hence we assume that the galaxies in our model evolve with
\begin{eqnarray}
  R(z) & = & \frac{R_0}{z+1}, \label{eq_R} \\ 
  H(z) & = & \frac{H_0}{z+1}. \label{eq_H}
\end{eqnarray}
Thus, the total volume of the disks
\begin{eqnarray}
  V(z) = \pi R(z)^2  H(z) 
\label{eq_Vgal}
\end{eqnarray}
is proportional to $(z+1)^{-3}$. \\
With the assumption that mass is conserved during the galaxy evolution, the particle density of the neutral gas\footnote{We point out that particle density denoted by $n$ always refers to the neutral gas and not the cosmic ray particles.} $n$ scales as $(z+1)^{3}$. We note that mass conservation is a further simplifying assumption. On the one hand galaxies gain mass through mergers and accretion. On the other hand they again loose material via galactic winds that will be discussed in the following. We assume here that these two effects do not change the total mass of a high-mass galaxy significantly during the phase of active star formation which is the main phase considered in this work. In addition to the redshift dependence of $n$ we also consider an increase of the gas density with the star formation rate density $\dot{\rho}_\star = \dot{M}_\star/V$. According to the empirical Kennicutt-Schmidt relation \citep{Kennicutt1998} we expect $\dot{\rho}_\star \propto n^{1.4}$. In our model the particle density scales as
\begin{eqnarray}
  n(z,\dot{\rho}_\star) = n_0~(z+1)^{3-\alpha} \left(\frac{\dot{\rho}_\star}{\dot{\rho}_{\star,0}}\right)^{1/1.4},
\label{eq_n}
\end{eqnarray}
where $n_0$ is the gas density and $\dot{\rho}_{\star,0}$ is the star formation rate density of our fiduical model at $z=0$. Due to the dependency on the galactic volume (\ref{eq_Vgal}) $\dot{\rho}_\star$ scales with $(1+z)^3$ for a fixed star formation rate. This leads to a scaling of the density of approximately $n\propto(1+z)^{5.14-\alpha}$. In order to test the influence of the gas density we include here an additional free parameter $\alpha$. For $\alpha=0$ the redshift dependence of $n$ is strong. In addition, we present results for $\alpha=2.14$. The latter corresponds to a case where $n \propto (z+1)^{3} (\dot{M}_\star/\dot{M}_{\star,0})^{1/1.4}$. \\
The present-day density values $n_0$ that we have chosen are listed in table \ref{Table_Props} and we present the redshift evolution of $n$ in figure \ref{plot_n_z}. We note that the gas in the interstellar medium can change from an atomic to a molecular phase which affects the number density. Observations show that the gas is mostly molecular above column densities of approximately $10^{21}$~cm$^{-2}$ \citep{LeroyEtAl2008,KennicuttEvans2012}. For hydrogen the number density can thus change by up to a factor of two. As we are performing order-of-magnitude estimates for the model galaxies, this factor is not significant.
\begin{table}
\begin{center}
\caption{\label{Table_Props}Properties of our fiducial galaxies}
     \begin{tabular}{lll}
      \hline  \hline  
      ~ 					& normal disk		& high-$z$ starburst \\
      \hline
       $R_0$~[pc] 				& $15000$  		& $3300$	\\
       $H_0$~[pc] 				& $500$  		& $165$		\\
       $n_0$~[cm$^{-3}$]        		& $3$ 			& $28$  	 \\
       $\dot{M}_{\star,0}$~[M$_\odot$~yr$^{-1}$]& $2$   		& $400$   	\\
       $v_\mathrm{wind,0}$~[km~s$^{-1}$]     	& $50$   		& $230$		\\
       $B_\mathrm{0}$~[$\mu$G]     		& $10$   		& $93$		\\
      \hline  \hline  
    \end{tabular}
\end{center}
\tablecomments{Summary of the properties of our fiducial galaxies: a normal disk galaxy comparable to the Milky Way and a starburst galaxy at high redshift comparable to SMM J2135-0102. We list the parameters that we employ at a redshift $z=0$. Listed are the galactic radius and scale height $R_0$ and $H_0$, the particle density $n_0$, the star formation rate $\dot{M}_{\star,0}$, the velocity of the galactic wind $v_\mathrm{wind,0}$, and the magnetic field strength $B_\mathrm{0}$. For the high-$z$ model we extrapolate the parameters observed at $z=2.3$ to $z=0$, i.e.~$R_0 = R_{2.3} (1+2.3)$, $H_0 = H_{2.3} (1+2.3)$, and $n_0 = R_{2.3} (1+2.3)^{-3+\alpha}$ with $R_{2.3} = 1000$ pc, $H_{2.3} = 50$, and $n_{2.3} = 1000$ cm$^{-3}$ (see section \ref{subsec_GeneralProperties}).}
\end{table}

\paragraph{Star formation and supernovae}

Important ingredients in our model are the star formation rate $\dot{M}_\star$ and the star formation rate density $\dot{\rho}_\star$. We use fixed values of $\dot{M}_{\star,0}$ at $z=0$ for our two fiducial galaxy models, the Milky Way and SMM J2135-0102, which are given in in table \ref{Table_Props}. In addition we perform calculations in which $\dot{M}_\star$ is treated as a free parameter and varied over a large range. \\
Stars with masses above roughly 8 solar masses ($M_\odot$) are determined to explode as supernovae. Assuming a Kroupa stellar initial mass function \citep{Kroupa2002} the number of supernovae per time is approximately 
\begin{eqnarray}
  \dot{N}_\mathrm{SN} = 0.156~\frac{\dot{M}_\star}{\overline{M}_\mathrm{SN}}
\label{eq_NSN}
\end{eqnarray}
with a mean mass of $\overline{M}_\mathrm{SN} \approx 12.26~M_\odot$.

\begin{table*}
\begin{center}
\caption{\label{Table_ISRF}The components of the thermal interstellar radiation field}
     \begin{tabular}{lllllll}
      \hline  \hline  
       ~ & ~ 		     		& UV 		      & optical 	    &  IR (warm) & IR (cold) 	      & CMB           \\
      \cline{2-7}
      normal disk:    	& $f_i$  	& $8.4\times10^{-17} \dot{\rho}_\star/\dot{\rho}_{\star,0}$ & $8.9\times10^{-13} \dot{\rho}_\star/\dot{\rho}_{\star,0}$ &	- 	 & $1.3\times10^{-5} \dot{\rho}_\star/\dot{\rho}_{\star,0}$ &  1           \\
        ~        	& $T_i$ [K] ~  	& $1.8\times10^{4}$   & $3.5\times10^{3}$   & 	-	 & $41$  	      & $2.73(1+z)$        \\
      \cline{2-7}
      high-$z$ starburst:  & $f_i$     	& $3.2\times10^{-16} \dot{\rho}_\star/\dot{\rho}_{\star,0}$ & -		    & $3.61\times10^{-6} \dot{\rho}_\star/\dot{\rho}_{\star,0}$ & $4.22\times10^{-2} \dot{\rho}_\star/\dot{\rho}_{\star,0}$  & 1   \\
      	~        	& $T_i$ [K] ~~ 	& $1.8\times10^{4}$   &	-   &	$60$	 & $30$       	      & $2.73 (1+z)$ \\
      \hline  \hline  
    \end{tabular}
\end{center}
\tablecomments{A model of the thermal interstellar radiation field which includes five different radiation components: ultraviolet (UV) radiation, optical radiation, thermal (warm and cold) infrared (IR) radiation, and the cosmic microwave background (CMB) (see \cite{CirelliPanci2009} and \cite{ChakrabortyFields2013}). We give here the dimensionless weights compared to the CMB $f_i$ which include a scaling with the normalized star formation rate density $\dot{\rho}_\star/\dot{\rho}_{\star,0}$ and the temperatures $T_i$. The fractions $f_i$ for the high-$z$ starburst are estimates from figure 1 in \citet{IvisonEtAl2010}.}
\end{table*}
\paragraph{Thermal interstellar radiation field}

The thermal part of the interstellar radiation field in our model is composed of the following components: the cosmic microwave background (CMB), a cold and a warm infrared (IR) component, an optical (opt) component, and an ultraviolet (UV) component \citep{CirelliPanci2009}. For the normal disk galaxy model we use the temperatures $T_i$ and weighting factors $f_i$ proposed in \citet{ChakrabortyFields2013}. Observational data of the interstellar radiation field is available for the high-$z$ galaxy. \citet{IvisonEtAl2010} present the rest-frame near-IR-radio SED in their figure 1. Besides the IR temperatures which are explicitly given in this study we extract the approximate radiation field parameters from their figure. While \citet{ChakrabortyFields2013} do not report an optical component of the interstellar radiation field, other starburst galaxies show emission in the optical (see e.g.~the spectrum of M82 in figure 17 in \citet{GallianoEtAl2008}). We have tested models with optical components ($T_\mathrm{opt}=3.5\times10^{3}$ K) for the starburst case and did not see any significant changes below a strength of $f_\mathrm{opt}\approx 10^{-10}\dot{\rho}_\star/\dot{\rho}_{\star,0}$. A summary of our radiation field model is given in table \ref{Table_ISRF}. \\
Assuming that the thermal radiation field is composed of various Planck components we calculate the total total energy density as
\begin{eqnarray}
  u_{\mathrm{ISRF}}(\dot{\rho}_\star)  & = & \int_0^\infty \left( \sum_i f_i(\dot{\rho}_\star)~\frac{8 \pi h}{c^3}~\frac{\nu^3}{\mathrm{exp}(h\nu/(k T_i) - 1} \right) ~\mathrm{d}\nu \nonumber \\
                     & = & \frac{8~\pi^5 k^4}{15~c^3 h^3}~\sum_i f_i(\dot{\rho}_\star) T_i^4
\label{uISRFtot}
\end{eqnarray}
with $i \in \{\mathrm{IR,opt,UV,CMB}\}$. We integrate here over the frequency $\nu$. Further, $c$ and $h$ are the speed of light and the Planck constant, respectively.\\
We emphasize that in our model all components besides for the CMB are scaling with the star formation rate density $\dot{\rho}_\star$. \citet{LackiEtAl2010b} on the other hand choose a dependence on the star formation rate surface density $\dot{\Sigma}_\star$.

\paragraph{Galactic winds}

Another important property of galaxies, especially in the context of star formation and cosmic rays, are galactic winds. The interplay of these three components has been studied in recent state-of-art numerical simulations \citep{GirichidisEtAl2015}. \\
With a higher star formation rate we expect more supernova explosions which drive the galactic outflows. Motivated from the analytical model of \citet{ShuMoShu-DeMao2005} we employ the following expression for the wind velocity
\begin{equation}
  v_\mathrm{wind}(\dot{\rho}_\star) = v_\mathrm{wind,0} \left(\frac{\dot{\rho}_\star}{\dot{\rho}_{\star,0}}\right)^{0.146}.
\label{eq_vwind}
\end{equation}
Outflow velocities of the Milky Way can be studied using numerical simulations. A recent study of a highly complex setup of a galactic disk by \citet{GirichidisEtAl2016} show that the bulk of the outflow occurs at low velocities of $20-40~\mathrm{km~s}^{-1}$. However they also observe a high velocity tail with a few $100~\mathrm{km~s}^{-1}$. We choose for our Milky Way model a value of $v_\mathrm{wind,0} = 50~\mathrm{km~s}^{-1}$. In case of the high-$z$ galaxy no observational data for the outflows is available. We thus use the wind velocity observed in M 82 which has a value of $230~\mathrm{km~s}^{-1}$ \citep{WalterEtAl2002} as a reasonable estimate.

\subsection{Magnetic fields}

\begin{figure}
  \includegraphics[width=0.5\textwidth]{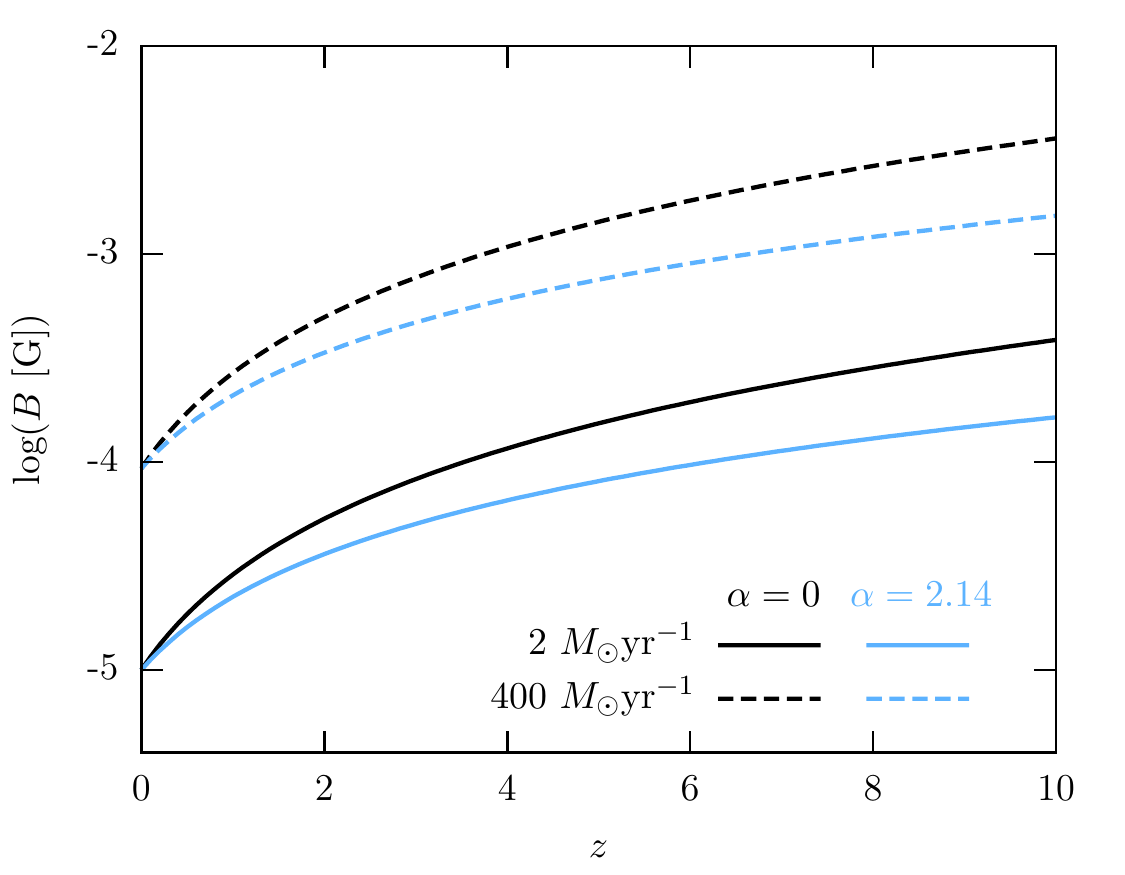}
  \caption{The magnetic field strength $B$ as a function of redshift $z$. As in figure \ref{plot_n_z}, the different linestyles show models with different star formation rates and different colors models with different density evolution (see equation \ref{eq_n}).}
\label{plot_B_z}
\end{figure}

A central assumption of this paper is that magnetic fields are omnipresent and strong in typical galaxies - also at high redshift. This assumption is based on the presence of a small-scale turbulent dynamo that can produce strong fields on short timescales by randomly stretching, twisting, and folding the magnetic field lines in turbulent motions \citep{Kazantsev1968,KulsrudAnderson1992,BrandenburgSubramanian2005,FederrathEtAl2011b,SchoberEtAl2012.1,SchoberEtAl2012.3,FederrathEtAl2014b}. Recently, the role of this turbulent dynamo has been analyzed in the formation of the first stars \citep{SchoberEtAl2012.2} and galaxies \citep{SchoberEtAl2013,LatifEtAl2013,PakmorSpringel2013}. A common prediction of these studies is that dynamo amplification leads to a certain fraction of equipartition between the turbulent and the magnetic energy. This fraction can be up to roughly 40 percent assuming ideal conditions \citep{FederrathEtAl2014b, SchoberEtAl2015}. The amplification timescale is short  compared to the dynamical timescales of a galaxy. \\
As mentioned above, the energy source of the dynamo is turbulence. The global turbulence content of a galaxy is driven by stellar feedback, most prominently by supernova shocks propagating through the interstellar medium \citep{MacLowKlessen2004,Gazol-PatinoPassot1999,KorpiBrandenburgTouminen1998}, and to some degree by accretion onto the disk \citep{KlessenHennebelle2010}. Both processes are directly related to the star formation rate, which is obvious for supernovae feedback and requires the assumption of steady state for accretion driven turbulence \citep{KlessenGlover2016}. In our model, the injection of turbulent kinetic energy is thus related to $\dot{M}_\star$ and the scale height of the galaxy $H$ which sets the timescale for turbulence decay. Under steady state the turbulent velocity $v_\mathrm{turb}$ can be estimated from equipartition between the loss rate $1/2 \rho v_\mathrm{turb}^2/(H/v_\mathrm{turb})$ and the energy injection rate $\dot{\rho_\mathrm{SN}} f_\mathrm{SN} E_\mathrm{SN}$. Here $\rho=n m$ is the mass density for particle masses $m$, $f_\mathrm{SN} E_\mathrm{SN}$ is the fraction of supernova energy converted into turbulence, and $\dot{\rho}_\mathrm{SN} = \dot{N}_\mathrm{SN}/V$ the SN rate density. Solving for $v_\mathrm{turb}$ results in
\begin{equation}
  v_\mathrm{turb} = \left(2 \dot{\rho_\mathrm{SN}} f_\mathrm{SN} E_\mathrm{SN} H \rho^{-1}\right)^{1/3}.
\label{eq_vturb}
\end{equation}
A fraction $\epsilon$ of the turbulent kinetic energy $1/2 \rho v_\mathrm{turb}^2$ is converted into magnetic energy $B^2/(8\pi)$. This leads to a scaling of the resulting field strength $B \propto \rho^{1/2} v_\mathrm{turb}$ and thus to
\begin{equation}
  B(z,\dot{\rho}_\star) = B_0~n(z)^{1/6} \left(\dot\rho_\star H(z)\right)^{1/3},
\label{eq_B}
\end{equation}
where we use $\dot\rho_\mathrm{SN} \propto \dot\rho_\star$. In agreement with the derivation based on small-scale dynamo action above, observations of various types of galaxies suggest a scaling of approximately $B \propto \dot\Sigma_\star^{0.3}$ \citep{NiklasBeck1997,ChyzyEtAl2011}. \\
As a local magnetic field strength $B_0$ we use observational results for the Milky Way. Using synchrotron emission \citet{Beck2001} find a value of $6\pm2~\mu$G in the solar neighborhood and $10\pm3~\mu$G at a 3 kpc galactic radius. We choose $B_0 = 10~\mu$G for our fiducial low star formation rate model. There is no observational data for magnetic fields in galaxies at high redshift, although we expect them to be in equipartition due to turbulent dynamo amplification. We thus employ the relation (\ref{eq_B}) to extrapolate a field strength from the Milky Way field strength. With the density, the star formation rate, and the size of SMM J2135-0102 as given in table \ref{Table_Props} we find $B_0 = 93~\mu$G. Observations of local starburst galaxies justify this high value of the magnetic field strength. For instance, \citet{AdebahrEtAl2013} report a field strength of 98 $\mu$G in the core region of M 82. The redshift evolution of the magnetic field strength that we employ for our model is presented in figure \ref{plot_B_z}.

\subsection{Cosmic rays}

\paragraph{Injection spectrum}
To study the physics of galaxies we use radiation emitted by cosmic rays. These high energy particles originate in shock fronts, e.g.~within supernova remnants, where thermal charged particles gain energy due to first-order Fermi acceleration \citep{Bell1978a,Bell1978b,Drury1983,Schlickeiser2002}. Theoretical studies of this acceleration mechanism predict the following injection spectrum of cosmic rays:
\begin{equation}
  Q(\gamma) = Q_{0}~\gamma^{-\chi},
\label{eq_Q}
\end{equation}
where $Q_{0}$ is the normalization, $\gamma = E /(m c^2)$ the Lorentz factor for a particle with energy $E$ and rest mass $m$, and the slope $\chi$ varies between $\chi=2.0$ for non-relativistic gas and $\chi=2.5$ for a relativistic gas \citep{Bell1978b}. We note that detailed models of supernova shock fronts result in $\chi = 2.1 - 2.3$ \citep{BogdanVolk1983} and choose a value of $\chi=2.2$ for our current study. \\
If we assume that galactic cosmic rays originate exclusively from supernova remnants, the total energy injection rate is $\xi E_\mathrm{SN} \dot{N}_\mathrm{SN}$, where $E_\mathrm{SN}$ is the energy of one supernova and $\dot{N}_\mathrm{SN}$ the supernova rate (\ref{eq_NSN}). The fraction of supernova energy that is converted into kinetic energy of cosmic rays $\xi$ can vary slightly with the density of the interstellar medium \citep{Dorfi2000}. We employ a fixed fiducial value of $\xi=0.1$. Due to their higher mass compared to electrons the energy is mostly within protons and ions. With a normalization to $\dot{N}_\mathrm{SN}$ we find for the proton injection spectrum
\begin{equation}
  Q(\gamma_\mathrm{p}) = Q_{\mathrm{p},0}~\gamma_\mathrm{p}^{-\chi}
\label{eq_Qp}
\end{equation}
the following proportionality constant:
\begin{equation}
  Q_{\mathrm{p},0} = \frac{\xi E_\mathrm{SN} \dot{N}_\mathrm{SN} (\chi-2)}{m_\mathrm{p} c^2 ~ \gamma_\mathrm{p,0}^{2-\chi}}.
\label{Qp0}
\end{equation}
As the upper end of the cosmic ray energy spectrum which extends up to $10^{21}$ eV per particle does not contribute significantly to the total cosmic ray energy, only the lower end of the spectrum $\gamma_\mathrm{p,0}$ appears here. For the latter we use a value of $\gamma_\mathrm{p,0} = 10^9~\mathrm{eV} / (m_\mathrm{p} c^2) \approx 1$ which corresponds to the proton rest mass.

\paragraph{Cosmic ray losses}
The injection spectrum can, however, differ significantly from the steady state spectrum of cosmic rays due to diffusion and loss processes. Cosmic ray diffusion is described by the following equation \citep{Longair2011}:
\begin{equation}
  \frac{\partial N(E)}{\partial t} = Q(E) + \frac{\mathrm{d}}{\mathrm{d} E} \left[b(E) N(E)\right] - \frac{N(E)}{\tau_\mathrm{cl}(E)} + D \nabla^2 N(E).
\label{DiffLossEq}
\end{equation}
The spectrum of cosmic rays $N$ depends on the injection spectrum $Q$ and the continuous energy losses given by the second term on the right hand side of (\ref{DiffLossEq}) which includes the cooling rate $b$. Catastrophic losses are included in the third term which depends on the catastrophic loss timescale $\tau_\mathrm{cl}$. Further, equation (\ref{DiffLossEq}) accounts for actual diffusion that is given by the last term with the diffusion constant $D$. For steady state the time derivative is zero. Under the assumption of spatial homogeneity also the diffusion term proportional to $\nabla^2 N$ can be neglected. Equation (\ref{DiffLossEq}) can then be approximated as
\begin{equation}
  0 = Q(E) + \frac{\mathrm{d}}{\mathrm{d} E} \left[b(E) N(E)\right] - \frac{N(E)}{\tau_\mathrm{cl}(E)}.
\label{DiffLossEq2}
\end{equation}
This equation is valid for cosmic ray protons as well as for electrons. However, for the different species different terms dominate. In the following we derive the steady-state spectrum of cosmic ray electrons from equation (\ref{DiffLossEq2}).

\paragraph{Energy losses of cosmic ray electrons}
\begin{figure*}
  \includegraphics[width=\textwidth]{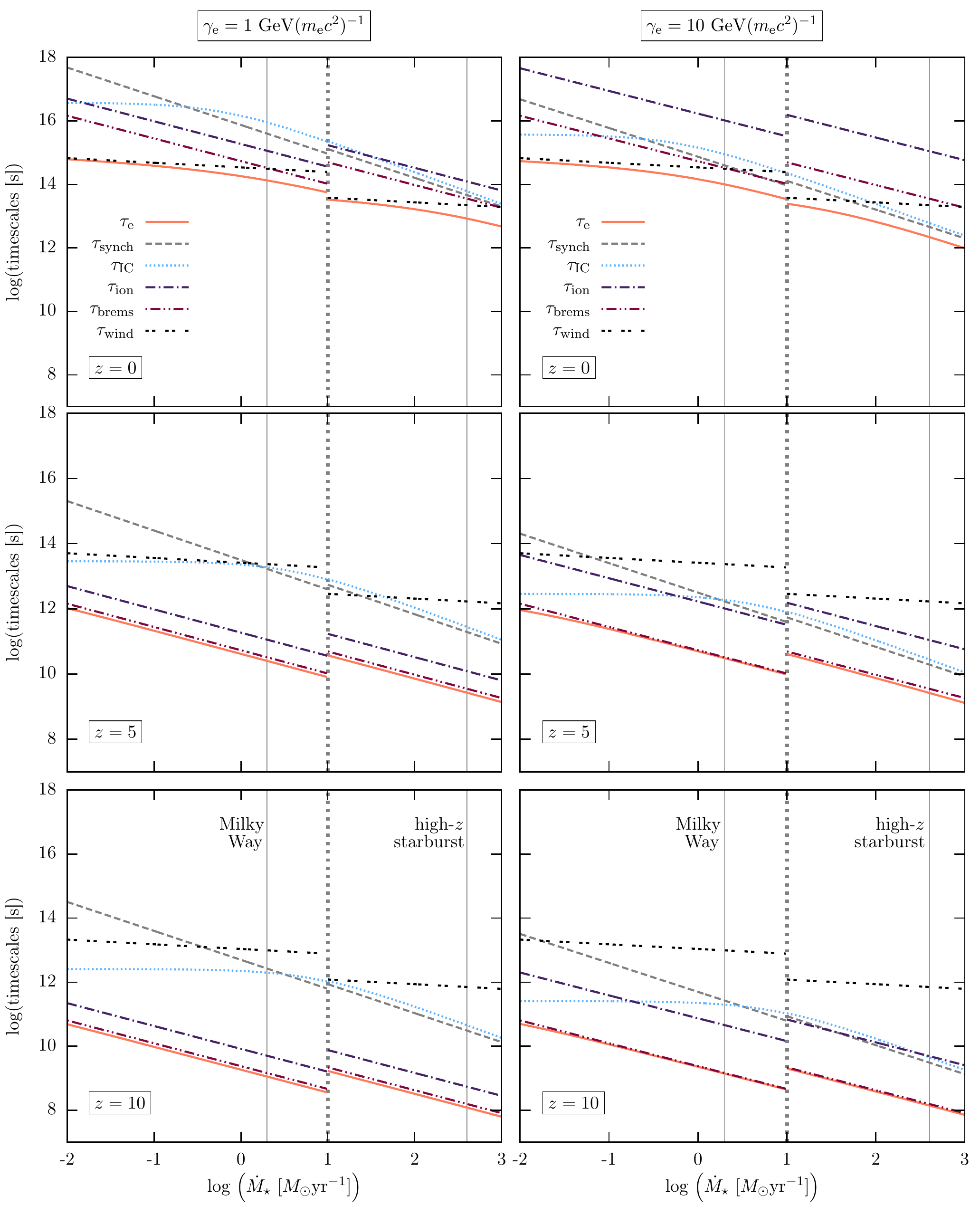}
  \caption{The energy loss timescale $\tau_\mathrm{e}$ and its individual components ($\tau_\mathrm{ion}$, $\tau_\mathrm{brems}$, $\tau_\mathrm{IC}$, $\tau_\mathrm{synch}$, $\tau_\mathrm{wind}$) of cosmic ray electrons with an energy of $\gamma_{e,0}$ (left panel) and $10~\gamma_{e,0}$ (right panel). We present the dependence on the star formation rate $\dot{M}_\star$. The different panels show different redshifts: $z=0$, $z=5$, $z=10$. The two different star formation regimes are separated by a thick vertical dashed line at $\dot{M}_\star = 10~M_\odot\mathrm{yr}^{-1}$. The positions of our fiducial galaxy models are marked by vertical thin lines: the Milky Way model at $\dot{M}_\star = 2~M_\odot\mathrm{yr}^{-1}$ and the high-$z$ starburst at $\dot{M}_\star = 400~M_\odot\mathrm{yr}^{-1}$. We chose here a density evolution with $\alpha=0$.}
\label{plot_timescales_SFR2}
\end{figure*}
High energy cosmic ray electrons loose energy mostly via cooling processes which are controlled by the cooling rate $b(E)$. The latter can be approximated by $E/\tau_\mathrm{e}(E)$ with $\tau_\mathrm{e}(E)$ being the cooling timescale. The following cooling mechanisms contribute significantly to $\tau_\mathrm{e}$: ionization ($\tau_\mathrm{ion}$), bremsstrahlung ($\tau_\mathrm{brems}$), inverse Compton scattering ($\tau_\mathrm{IC}$), synchrotron emission ($\tau_\mathrm{synch}$), and adiabatic losses ($\tau_\mathrm{wind}$). The total cooling timescale is calculated as
\begin{equation}
  \tau_\mathrm{e} = \left(\tau_\mathrm{ion}^{-1} + \tau_\mathrm{brems}^{-1} + \tau_\mathrm{IC}^{-1} + \tau_\mathrm{synch}^{-1} + \tau_\mathrm{wind}^{-1}\right)^{-1}.
\label{te}
\end{equation}
with the individual contributions
\begin{eqnarray}
  \tau_\mathrm{ion} & = & \frac{\gamma_\mathrm{e}}{2.7~c~\sigma_\mathrm{T}~(6.85 + 0.5~\mathrm{ln}\gamma_\mathrm{e})~n} \label{tau_ion}, \\
  \tau_\mathrm{brems} & = & 3.12 \times10^7~\mathrm{yr}~\left(\frac{n}{\mathrm{cm}^{-3}}\right)^{-1} \label{tau_brems}, \\
  \tau_\mathrm{IC} & = & \frac{3~m_\mathrm{e}~c}{4~\sigma_\mathrm{T}~u_\mathrm{ISRF}~\gamma_\mathrm{e}} \label{tau_IC}, \\
  \tau_\mathrm{synch} & = & \frac{3~m_\mathrm{e}~c}{4~\sigma_\mathrm{T}~u_\mathrm{B}~\gamma_\mathrm{e}} \label{tau_synch}, \\
  \tau_\mathrm{wind} & = & \frac{H}{v_\mathrm{wind}}. \label{tau_wind}  
\end{eqnarray}
For the ionization timescale we refer to \citet{Schlickeiser2002} and for the bremsstrahlung timescale to \citet{StrongMoskalenko1998}. For both timescales we assume that the particle density of hydrogen is ten times the one of Helium. We further use the density of the interstellar radiation field $u_\mathrm{ISRF}$ given in (\ref{uISRFtot}) and the magnetic energy density $u_\mathrm{B} = B^2/(8 \pi)$ with the field strength given in (\ref{eq_B}). The wind velocity $v_\mathrm{wind}$ depends on the star formation rate and the scale height of the galaxy and is given in equation (\ref{eq_vwind}). \\
We present the cooling timescale of cosmic ray electrons as a function of the star formation rate in figure \ref{plot_timescales_SFR2}. The timescales shown in the left panel are calculated for cosmic ray electrons with a Lorentz factor of $\gamma_\mathrm{e} = \gamma_\mathrm{e,0} = 10^7~\mathrm{eV} / (m_\mathrm{e} c^2)$, while the energy is ten times higher in the right panel. In particular we stress that the timescales for synchrotron emission and inverse Compton scattering decrease with increasing cosmic ray energy in the same way as both are inversely proportional to $\gamma_\mathrm{e}$. The two regimes, the normal disk regime for $\dot{M}_\star < 10~M_\odot\mathrm{yr}^{-1}$ and the starburst regime for $\dot{M}_\star > 10~M_\odot\mathrm{yr}^{-1}$, are separated by a vertical dashed line. The positions of the fiducial galaxy models are marked by the thin vertical lines. The redshift increases from the upper to the lower panels. Note that all the individual components of $\tau_\mathrm{e}$ decrease with redshift. One of the most significant effects is the decrease of $\tau_\mathrm{ion}$ and $\tau_\mathrm{brems}$ with $z$ which origin in their dependence on the density $n$. 
\begin{figure}[t]
  \includegraphics[width=0.5\textwidth]{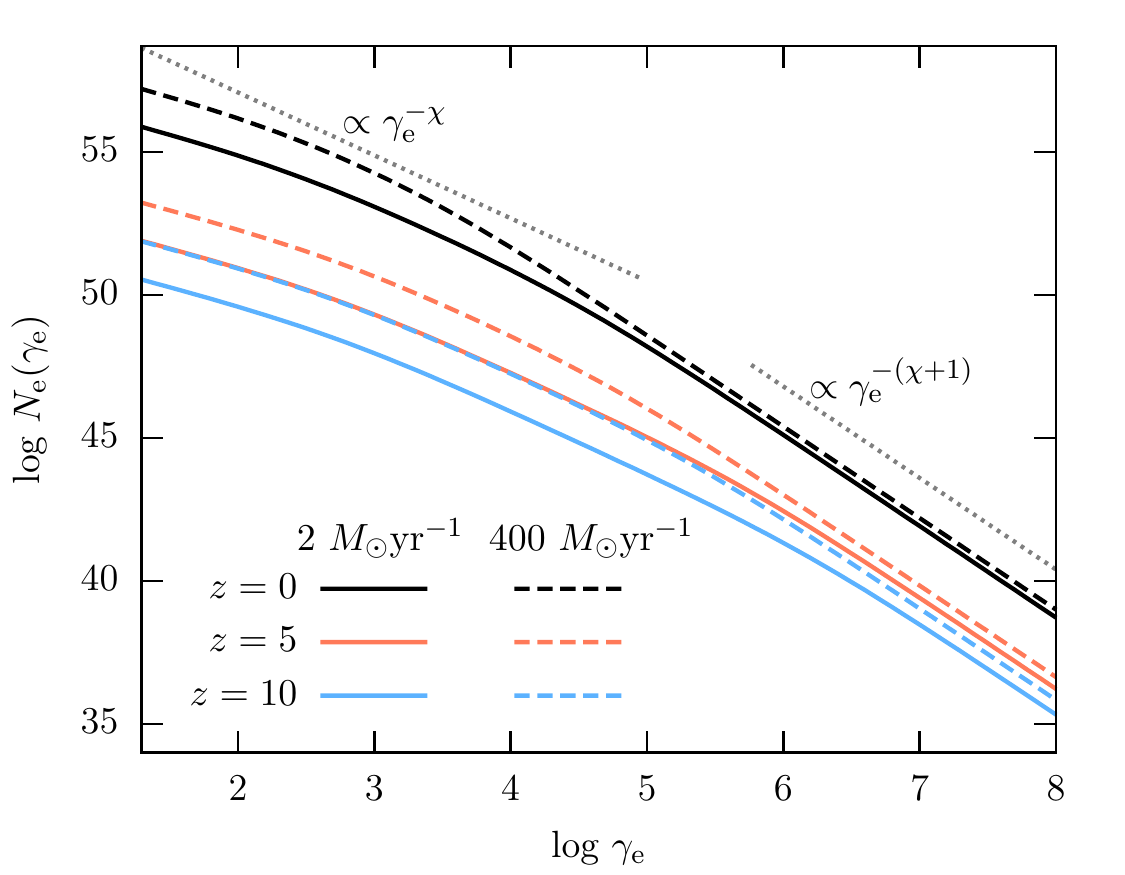}
  \caption{The number of cosmic ray electrons $N_e$ as a function of the Lorentz factor $\gamma_e$ in steady state. The solid lines show the results for the Milky Way model with $\dot{M}_\star=2~M_\odot\mathrm{yr}^{-1}$, while the dashed lines show the ones for the high-$z$ starburst galaxy with $\dot{M}_\star=400~M_\odot\mathrm{yr}^{-1}$. Different colors indicate different redshifts $z$. For reference we indicate the scaling of the injection spectrum which is proportional to $\gamma_\mathrm{e}^{-\chi}$ and the scaling $\gamma_\mathrm{e}^{-(\chi+1)}$ which is expected at high $\gamma_\mathrm{e}$ by the dashed lines. We chose here a density evolution with $\alpha=0$.}
\label{plot_Gamma_Ne2}
\end{figure}
The latter increases proportional to $(1+z)^3$. As a result the total cosmic ray cooling timescale at high redshift scales as the density with $\dot{\rho}_\star^{1/1.4}$. \\
In contrast to the cosmic ray protons, the electrons do not decay. Thus there are no catastrophic losses of $e^\pm$ and we can neglect the last term of equation (\ref{DiffLossEq2}). The steady state spectrum for $e^\pm$ is then given by:
\begin{equation}
  N_\mathrm{e}(\gamma_\mathrm{e}) = \frac{Q_\mathrm{e}(\gamma_\mathrm{e})~\tau_\mathrm{e}(\gamma_\mathrm{e})}{\chi - 1}.
\label{eq_Ne1}
\end{equation}

\paragraph{Contribution from secondary cosmic ray electrons}
The total energy injection rate $Q_\mathrm{e}$ of cosmic ray $e^\pm$ is composed of a primary component $Q_{\mathrm{e},0}$ from supernova shocks and a secondary component $Q_\mathrm{e,sec}$ from the decay of proton produced pions. In fact, gamma-ray observations suggest that the secondary electrons dominate with a fraction of roughly 70 percent \citep{LackiEtAl2011}. We label the ratio of the secondary injection rate $Q_\mathrm{e,sec}$ to the total injection rate $Q_\mathrm{e} = Q_{\mathrm{e},0} + Q_\mathrm{e,sec}$ as $f_\mathrm{sec} = Q_\mathrm{e,sec} / Q_\mathrm{e} \approx 0.7$. \\
For high energies cosmic ray protons experience mostly catastrophic losses via pion production \citep{MannheimSchlickeiser1994} the timescale for which is $\tau_\mathrm{cl,p} = \tau_\pi f_\pi$. Here $\tau_\pi$ is the characteristic time for pion production and $f_\pi$ the fraction of protons contributing to that. For the latter fraction we assume a value of $f_\pi=0.4$ \citep{LackiBeck2013}. Neglecting continuous losses, i.e.~setting the second term on the right hand side of equation (\ref{DiffLossEq2}) to zero, one finds for the steady state spectrum of protons
\begin{equation}
  N_\mathrm{p}(\gamma_\mathrm{p}) = f_\pi \tau_\pi Q_{\mathrm{p},0}~\gamma_\mathrm{p}^{-\chi} m_\mathrm{p} c^2,
\end{equation}
where the additional factor of $m_\mathrm{p} c^2$ comes from the conversion from $E_{\mathrm{p}}$ to $\gamma_{\mathrm{p}}$. \\
\citet{LackiBeck2013} derive the following relation between the secondary $e^\pm$ and the injection rate of cosmic ray protons $Q_\mathrm{p}$:
\begin{equation}
  Q_\mathrm{e,sec}(E_\mathrm{e,sec}) = \frac{f_\pi}{6} \left(\frac{E_\mathrm{p}}{E_\mathrm{e,sec}}\right)^2 Q_\mathrm{p}(E_{\mathrm{p}}),
\end{equation}
where they assume for the energy of the protons $E_\mathrm{p}= 20 E_\mathrm{e,sec}$. In terms of the Lorentz factor $\gamma_\mathrm{e}= E_\mathrm{e}/(m_\mathrm{e} c^2)$ we get
\begin{equation}
  Q_\mathrm{e,sec}(\gamma_\mathrm{e}) = \frac{20^{2-\chi}}{6} f_\pi \left(\frac{m_\mathrm{p}}{m_\mathrm{e}}\right)^{\chi} Q_\mathrm{p,0} \gamma_\mathrm{e}^{-\chi}~ m_\mathrm{e} c^2.
\label{Qesec}
\end{equation}

\paragraph{Steady-state spectrum of cosmic ray electrons}
Taking into account the injection of secondary $e^\pm$ we find the following steady state spectrum from equation (\ref{eq_Ne1}):
\begin{eqnarray}
  N_\mathrm{e}(\gamma_\mathrm{e}) & = & \frac{Q_\mathrm{sec, e}(\gamma_\mathrm{e})~\tau_\mathrm{e}(\gamma_\mathrm{e})}{f_\mathrm{sec}(\chi - 1)} \nonumber \\
 				  & = & \frac{20^{2-\chi}}{6 (\chi-1)} \frac{f_\pi}{f_\mathrm{sec}} \left(\frac{m_\mathrm{p}}{m_\mathrm{e}}\right)^{\chi} \tau_\mathrm{e}(\gamma_\mathrm{e})~Q_\mathrm{p,0} \gamma_\mathrm{e}^{-\chi}~m_\mathrm{e} c^2. \nonumber \\
\label{Ne}
\end{eqnarray}
The resulting steady state spectra for the two fiducial models at different redshifts are presented in figure \ref{plot_Gamma_Ne2}. Due to the decreasing energy loss timescale the number of cosmic ray electrons decreases with $z$. Note, that the slope of the steady state spectrum is steeper than the one of the injection spectrum which is indicated by the dotted line for reference. This effect is strongest for high energies $\gamma_\mathrm{e}$ and for galaxies with higher star formation rates. In these cases the cosmic ray energy losses are more significant. In fact, the scaling of the steady-state spectrum at large $\gamma_\mathrm{e}$ can be estimated from equation (\ref{DiffLossEq2}) which then takes the form of $0 = Q(\gamma_\mathrm{e}) + \frac{\mathrm{d}}{\mathrm{d} \gamma_\mathrm{e}} \left[\gamma_\mathrm{e}^2 N(\gamma_\mathrm{e})\right]$ except for constant factors. The factor $\gamma_\mathrm{e}^2$ results from the cooling timescale which is inversely proportional to $\gamma_\mathrm{e}$ at high energies. With $Q(\gamma_\mathrm{e})\propto\gamma_\mathrm{e}^{-\chi}$ we find a steady-state spectrum $N(\gamma_\mathrm{e})\propto\gamma_\mathrm{e}^{-(\chi+1)}$ for $\gamma_\mathrm{e}\gg1$. We indicate this scaling in figure \ref{plot_Gamma_Ne2}.



\section{Galactic radio emission}
\label{sec_Synchrotron}

\subsection{Non-thermal radio emission}

\begin{figure*}
  \includegraphics[width=\textwidth]{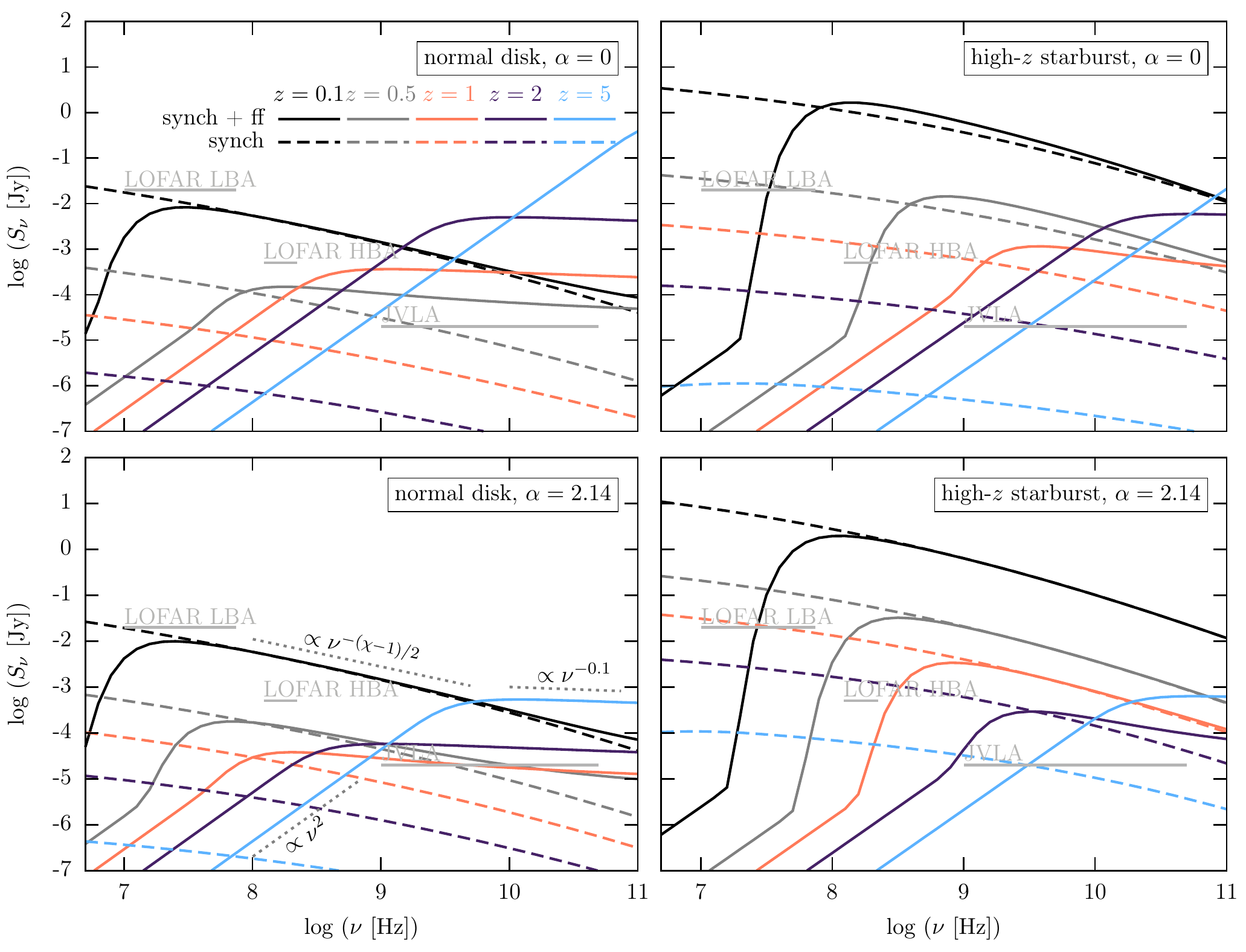}
  \caption{The observed spectral fluxes for the Milky Way-like model with a star formation rate of $\dot{M}_\star = 2~M_\odot\mathrm{yr}^{-1}$ (left panels) and for a typical high-$z$ starburst galaxy with $\dot{M}_\star = 400~M_\odot\mathrm{yr}^{-1}$ (right panels). In the top panels we employ a strong density evolution with redshift, e.g.~$\alpha=0$, while $\alpha=2.14$ in the lower panels (see equation \ref{eq_n} for the definition of $\alpha$). The dashed lines represent the pure non-thermal spectra as given in equation (\ref{eq_Snu}). The solid lines include additional free-free emission and absorption (see equation \ref{eq_totflux}). In the lower left panel we present fits for different scalings as discussed in section \ref{sec_ObsLimits}. Different colors indicate different redshifts $z$. Conservative observational limits for current radio telescopes are given as vertical gray lines.}
\label{plot_Snu_nu}
\end{figure*}

In a magnetic field a charged particle performs spiral motions around the field lines. Thus, it is constantly accelerated leading to synchrotron emission. The emitted energy from an electron with relativistic velocity $\boldsymbol{\beta}$ per time $t$ and unit solid angle $\Omega_n$ into the direction $\mathbf{n}$ is 
\begin{eqnarray}
  \frac{\mathrm{d}L_{\gamma_\mathrm{e}}}{\mathrm{d}\Omega_n} = \frac{e^2}{4\pi~c}~\frac{\left|\mathbf{n} \times \left[\left(\mathbf{n} - \boldsymbol{\beta}\right) \times \dot{\boldsymbol{\beta}}\right]\right|^2}{\left(1-\mathbf{n}\cdot\boldsymbol{\beta}\right)^6}.
\end{eqnarray}
It can be shown (see e.g.~the review by \citet{BlumenthalGould1970}) that the resulting spectral power from a single electron with energy $\gamma_e$ is
\begin{eqnarray}
  L_{\nu,\gamma_\mathrm{e}}(\nu, z, \dot{M}_\star, \gamma_\mathrm{e}) = \frac{\sqrt{3}~e^3~ B}{m_e c^2}~\frac{\nu}{\nu_\mathrm{c}}~ \int_{\nu/\nu_\mathrm{c}}^\infty K_{5/3}(x)~\mathrm{d}x, \nonumber \\
\label{eq_singlesynch}
\end{eqnarray}
where $K_{5/3}(x)$ is the modified Bessel function of second kind. This spectrum peaks roughly at the critical frequency 
\begin{eqnarray}
  \nu_\mathrm{c}(z,\dot{M}_\star,\gamma_\mathrm{e}) = \frac{3 \gamma_\mathrm{e}^2~e~B(z, \dot{M}_\star)}{4\pi~c~m_\mathrm{e}}
\end{eqnarray}
which is a function of the star formation rate that enters via the magnetic field and the cosmic ray energy. In our model we consider $\dot{M}_\star$ between $10^{-2}$ and $10^3~M_\odot \mathrm{yr}^{-1}$. For Lorentz factors of $\gamma_{e,0}=1~\mathrm{GeV}/(m_\mathrm{e} c^2)$ and $\gamma_{e,0}=10~\mathrm{GeV}/(m_\mathrm{e} c^2)$ and $z=0$ the critical frequency range of synchrotron emission is $1.5\times10^{7}-2.7\times10^{9}$ Hz and $1.5\times10^{9}-2.7\times10^{11}$ Hz, respectively. \\
For a distribution of electrons with different energies (\ref{Ne}), the spectral emission can then be determined by \citep{BlumenthalGould1970}
\begin{eqnarray}
  L_{\nu}(\nu, z, \dot{M}_\star) & = & \int_{\gamma_{\mathrm{e},0}}^\infty L_{\nu,\gamma_e}(\nu, z, \dot{M}_\star, \gamma_e) N_\mathrm{e}(\gamma_\mathrm{e}) ~\mathrm{d}\gamma_\mathrm{e} \nonumber \\
  & & \times \int N(\alpha) (\mathrm{sin}(\alpha))^{(\chi+1)/2}~\mathrm{d}\Omega_\alpha.
\label{eq_Lnu}
\end{eqnarray}
Here an integral over an pitch angel $\Omega_\alpha$ is included. For local isotropy the function $N(\alpha)$ equals 1 and one finds that the latter integral is roughly 8.9 for a cosmic ray distribution with an exponent of $\chi=2.2$. In case of a simple power law distribution with $N_\mathrm{e} \propto \gamma_\mathrm{e}^{-\chi}$ the integration of equation (\ref{eq_Lnu}) would yield $L_{\nu} \propto \nu^{-(\chi-1)/2}$. As we take into account various cooling processes for determining the cosmic ray spectrum $N_\mathrm{e}$ can, however, differ (see figure \ref{plot_Gamma_Ne2}) which has an impact on the resulting synchrotron spectrum.\\
The synchrotron flux observed from an object at a luminosity distance 
\begin{eqnarray}
  d_\mathrm{L}(z) = \frac{2 c}{H_0 \Omega_m^2}  \left(\Omega_m z+(\Omega_m-2) \left(\sqrt{\Omega_m z+1}-1\right)\right) \nonumber \\
\label{eq_dL}
\end{eqnarray}
is calculated by
\begin{eqnarray}
  S_{\nu,\mathrm{synch}}(\nu, z, \dot{M}_\star) = \frac{L_{\nu}(\nu, z, \dot{M}_\star)}{4 \pi~d_\mathrm{L}^2}.
\label{eq_Snu}
\end{eqnarray}
We use here a mass density parameter of $\Omega_m=0.32$ and a Hubble constant of $H_0=67~\mathrm{km~s}^{-1}\mathrm{Mpc}^{-1}$. The representative spectral fluxes calculated from our model of galactic synchrotron emission are shown in figure \ref{plot_Snu_nu}. Here we present the spectrum in the observed frame, including the observed frequencies. We again consider our fiducial galaxy models: the Milky Way based model is shown in the left panels and the high-$z$ starburst model in the right ones. Additionally, we show the results for a strong density evolution with redshift, i.e.~$\alpha=0$, and for a weak evolution, i.e.~$\alpha=2.14$ (see equation \ref{eq_n}). We present the pure non-thermal spectrum by dashed lines with different colors representing the fluxes at different redshifts. The synchrotron flux decreases with increasing redshift while the starburst lines lie always above the ones for the Milky Way galaxy. For a stronger density evolution, i.e.~$\alpha=0$, the flux decreases faster than for $\alpha=2.14$.

\subsection{Contribution from free-free emission}
\label{sec_freefree}
Especially at low frequencies free-free emission and absorption contribute to the radio spectrum and need to be included when designing observational studies. We note, however, that this contribution to the emission is not considered in the next section where we study the FIR-radio correlation. The free-free optical depth is given as
\begin{eqnarray}
  \tau_\mathrm{ff}(z, \dot{M}_\star) & = & 0.082\left(\frac{T_\mathrm{e}}{\mathrm{K}}\right)^{-1.35} \left(\frac{EM(z, \dot{M}_\star)}{\mathrm{cm^{-6}~pc}}\right)  \nonumber \\
   &   &  \times \left(\frac{\nu}{10^9~\mathrm{Hz}}\right)^{-2.1},
\label{eq_tauff}
\end{eqnarray}
where we assume an electron temperature $T_\mathrm{e}$ of $10^4$ K and an emission measure of
\begin{eqnarray}
  EM(z, \dot{M}_\star) \approx n_\mathrm{e}(z, \dot{M}_\star)^2~ H(z)~ f^{-1}.
\label{eq_EM}
\end{eqnarray}
Here we approximate the electron number density as $n_\mathrm{e}=0.1~n(z, \dot{M}_\star)$, i.e.~we employ an ionization degree of ten percent which is typical for the warm interstellar medium \citep{Tielens2005}. The propagation length through ionic material is estimated as the scale height of the galaxy (\ref{eq_H}) and we employ a filling factor of $f=0.3$ \citep{EhleBeck1993,BerkhuijsenEtAl2006,Beck2007}. We note that the optical depth increases strongly with redshift as the emission measure (\ref{eq_EM}) scales with $n^2$. With our expression for the density as given in (\ref{eq_n}) we find $\tau_\mathrm{ff}\propto(1+z)^{9.28-2\alpha}$ in the observed frequency frame. At an emitted frequency of $10^9$ Hz the transition from the optically thin to optically thick occurs at $z=1.9$ for the normal disk model with $\alpha=0$ and at $z=1.0$ for $\alpha=2.14$. In case of our fiducial starburst model $\tau_\mathrm{ff}=1$ at $z=1.9$ for $\alpha=0$ and at $z=0.5$ for $\alpha=2.14$.\\
Taking into account the effect of free-free absorption, the synchrotron emission is reduced by a factor of $\mathrm{e}^{-\tau_\mathrm{ff}}$. In addition, however, we expect also a positive contribution to the radio flux from free-free emission which can be calculated as
\begin{eqnarray}
  S_{\nu,\mathrm{ff}}(\nu,z, \dot{M}_\star) &=&  2~k~T_\mathrm{e}~c^{-2}~\Delta\Omega~(1 - \mathrm{e}^{-\tau_\mathrm{ff}(z, \dot{M}_\star)})~\nu^2, \nonumber\\
\label{eq_ff}
\end{eqnarray}
where we estimate the solid angle of a typical galaxy as $\Delta\Omega = \Delta A/ d_\mathrm{A}^2$. The observed surface area $\Delta A$ of a galaxy depends on its orientation with respect to the line of sight. For the model disk galaxies we assume here a mean value of $\Delta A= 0.5 \pi R(z)^2$. The angular diameter distance $d_\mathrm{A}$ can be calculated from the luminosity distance (\ref{eq_dL}) via $d_\mathrm{A} = (1+z)^{-2} d_\mathrm{L}$. We note that the factor $(1 - \mathrm{e}^{-\tau_\mathrm{ff})}$ becomes $1$ in the optical thick regime, i.e.~for $\tau_\mathrm{ff}\gg1$, while it reduces to $\tau_\mathrm{ff}$ in the optical thin regime. We thus expect the free-free flux density to scale with $\nu^2$ for $\tau_\mathrm{ff}\gg1$ and with $\nu^{-0.1}$ for $\tau_\mathrm{ff}\ll1$. \\
The total radio flux, including free-free absorption of the non-thermal flux and free-free emission, is given as
\begin{eqnarray}
  S_\nu = S_{\nu,\mathrm{synch}}~\mathrm{e}^{-\tau_\mathrm{ff}} + S_{\nu,\mathrm{ff}}.
\label{eq_totflux}
\end{eqnarray}
We present the total flux $S_\nu$ as solid lines in figure \ref{plot_Snu_nu}. It is clear from the figure that free-free absorption and emission play an important role for the total radio spectrum. As discussed above, $\tau_\mathrm{ff}$ and also the factor $(1-\mathrm{e}^{-\tau_\mathrm{ff}})$ increase strongly with $z$. Thus, even though the solid angle of the source $\Delta\Omega$ decreases monotonously with redshift, there is a regime at high frequencies and moderate redshifts in which $S_\mathrm{ff}$ increases with $z$. In our models of Milky Way like galaxies, the spectrum is dominated by free-free emission for $z\gtrsim 0.5$ for $\alpha=0$ and for $z\gtrsim 1$ for $\alpha=2.14$. For the starburst case the synchrotron emission dominates up to moderate redshifts as shown in the right panel of figure \ref{plot_Snu_nu}. At $z\gtrsim 1$ for $\alpha=0$ and $z\gtrsim 2$ for $\alpha=2.14$ synchrotron emission is, however, suppressed at low frequencies due to efficient free-free absorption.
\begin{table}
\begin{center}
\caption{\label{Table_telescopes}Sensitivities for different radio telescopes}
    \begin{tabular}{lll}
    \hline  \hline  
      	telescope	& frequency range	& sensitivity [mJy] \\
    \hline
	LOFAR LBA\footnote{http://www.lofar.org/} 	& 10-80 MHz 		& 20 \\
	LOFAR HBA 	& 120-240 MHz		& 0.5 \\
	JVLA\footnote{http://www.vla.nrao.edu/}		& 1-50 GHz 		& 0.02 \\
    \hline
    \hline
    \end{tabular}
\end{center}
\tablecomments{Frequency ranges and continuum sensitivities for an 8 hour exposure time of different radio telescopes. For the sensitivities are relatively conservative estimates for point sources, i.e.~at least 5 times rms noise.}
\end{table}

\subsection{Observational limits for radio telescopes}
\label{sec_ObsLimits}

For reference we indicate the limits for current radio telescopes in figure \ref{plot_Snu_nu}. The conservative estimates for the sensitivities and the corresponding frequency ranges are listed in table \ref{Table_telescopes}. Figure \ref{plot_Snu_nu} indicates that we can observe the non-thermal radio fluxes from galaxies up to a redshift of roughly 2 in case of a high star formation rate. \\
When comparing the models with real observational data, one needs to take into account the evolution of the free-free optical depth $\tau_\mathrm{ff}$. As our models show the latter becomes more important at higher redshifts. As a result synchrotron emission is absorbed at low frequencies and free-free emission dominates the spectrum at high $\nu$. This leads to a possible underestimate of the synchrotron flux at low $\nu$, while it can be overestimated at high $\nu$. Isolating the non-thermal contribution becomes more difficult at high redshift as the latter decreases with $z$ while free-free emission can increase at high frequencies. Multi-frequency observations are crucial to study the spectral slope and extract the non-thermal part of the spectrum. If the flux is proportional to $\nu^{-(\chi-1)/2}$ synchrotron emission dominates, while free-free emission dominates when $\nu^{-0.1}$ at high frequencies. In case of subdominant non-thermal emission, polarization measurements can be used to study the synchrotron contribution. From figure \ref{plot_Snu_nu} we conclude that the non-thermal contribution of the spectrum can be studied best for starburst galaxies. The optimal observational frequency is just above the cut-off through free-free absorption. In the observed frame as presented in figure \ref{plot_Snu_nu}, this frequency shifts from $\nu\approx10^8$ Hz to higher values\footnote{We note that typically one would expect a spectral feature to move to lower frequencies when increasing the redshift. Here, however, the peak of the spectrum moves to higher $\nu$ for higher redshifts as a result of the strong $z$ dependency of the free-free optical depth.}. While for \mbox{LOFAR} measurements in the low-$z$ regime are possible, the JVLA would be more suitable for observations at $z\gtrsim1$.\\
The future generation of radio telescopes will lower the detection limits further. For example the expected continuum sensitivity of the SKA precursor \textit{MeerKat}\footnote{http://www.ska.ac.za/meerkat/index.php} is roughly 1 $\mu$Jy for 8 hour integration. ASKAP\footnote{http://www.atnf.csiro.au/projects/askap/index.html} is a further SKA precursor with a sensitivity of the order of 10 $\mu$Jy and thus slightly below JVLA. SKA\footnote{https://www.skatelescope.org/} will be the most important future telescope to study radio emission from the early Universe. The SKA\textit{-mid} which will cover a frequency range of 0.35-14 GHz is expected to have a sensitivity of 0.3 $\mu$Jy for an exposure time of 8 hours. Based on the results of our theoretical model, non-thermal galactic radio emission from galaxies beyond $z=2$ should be easily detectable with the SKA. With a very long integration time of for example 100 hours, the sensitivity goes down to 0.07 $\mu$Jy and synchrotron emission from redshifts below 5 can be detected if the source galaxy is comparable to our fiducial high-$z$ galaxy SMM J2135-0102.



\section{FIR-radio correlation}
\label{sec_FIR-radio}

\subsection{In the local Universe}
\label{sec_FIR-radio_locally}

For calculating the far-infrared luminosity we employ a conversion relation given in \citet{Kennicutt1998} between $L_\mathrm{FIR}$, i.e.~IR luminosity integrated over the full-, mid-, and far-IR spectrum (wavelength band of 8-1000 $\mu$m), and the star formation rate $\dot{M}_\star$:
\begin{eqnarray}
  L_\mathrm{FIR}(\dot{M}_\star) = 5.79\times10^9~\frac{\dot{M}_\star}{M_\odot \mathrm{yr}^{-1}}~L_\odot.
\label{eq_LFIR}
\end{eqnarray}
We note that the relation in equation (\ref{eq_LFIR}) is only valid under a number of assumptions. First of all it was derived for starburst galaxies in which the FIR radiation field originates from the UV radiation of young stars heating up the dust. It is assumed that the dust reradiates the energy in the infrared. The study of \citet{Kennicutt1998} further uses the radiative transfer model of \citet{LeithererHeckman1995}, assumes a mean luminosity for 10-100 Myr continuous bursts, solar metal abundances, and a Salpeter IMF \citep{Salpeter1955}. We note that the FIR luminosities reported by \citet{YunEtAl2001} which we will discuss in the following result from observations of the 60 $\mu$m emission line. It has been argued that the monochromatic luminosity $L_{60~\mu\mathrm{m}}$ is related to the bolometric FIR luminosity by $L_\mathrm{FIR} \approx 1.7~L_{60~\mu\mathrm{m}}$ \citep{ChapmanEtAl2000}. 
\begin{figure}[t]
  \includegraphics[width=0.5\textwidth]{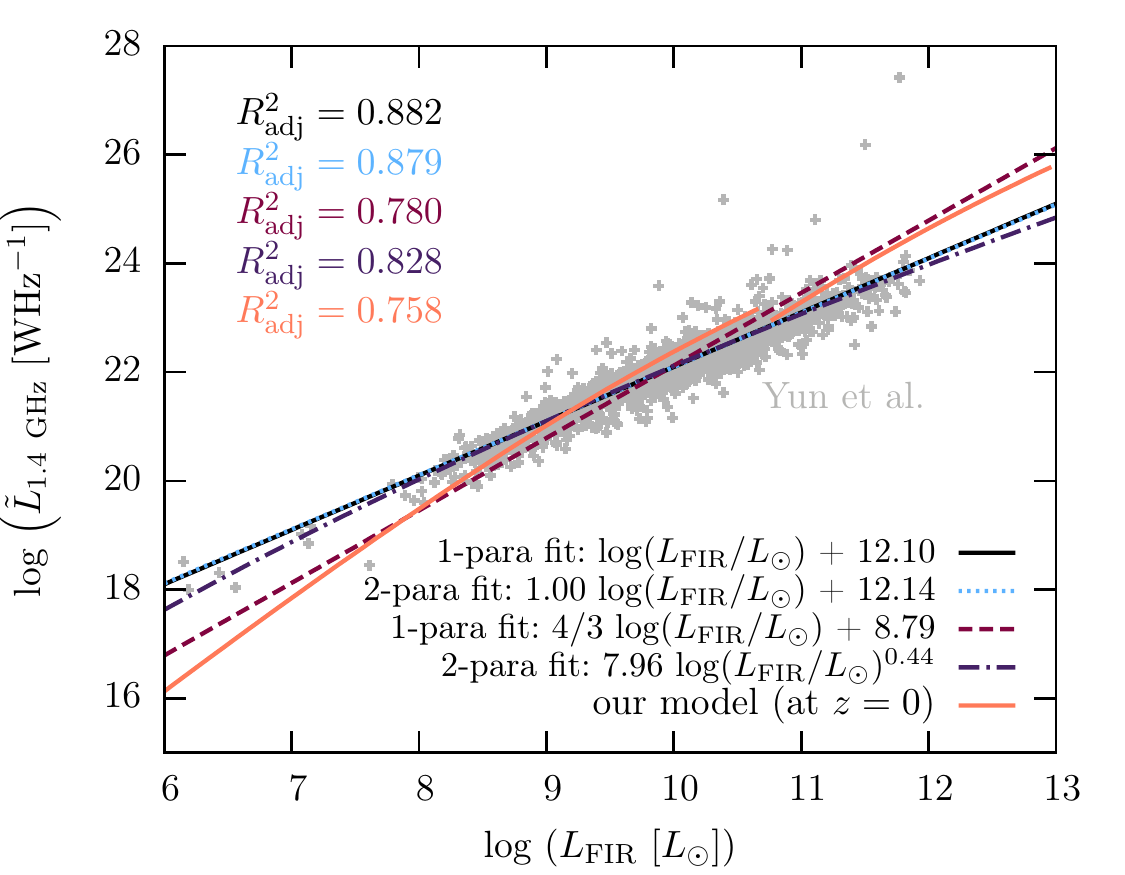}
  \caption{The FIR-radio correlation from the observational data of \citet{YunEtAl2001} given by the gray crosses. We overlay our theoretical model at $z=0$ with the orange line. In addition we perform various fits: $\mathrm{log}(L_\mathrm{FIR}/L_\odot) + c_1$, $c_2~\mathrm{log}(L_\mathrm{FIR}/L_\odot) + c_3$, $4/3~\mathrm{log}(L_\mathrm{FIR}/L_\odot) + c_4$, $c_5~\mathrm{log}(L_\mathrm{FIR}/L_\odot)^{c_6}$. The resulting fit parameters $c_i$ (for $i=1,..,6$) are given in the plot legend. For all curves (fits and model) we note the corresponding $R^2_\mathrm{adj}$ as introduced below in equation (\ref{eq_R2}) in the color of the corresponding line.}
\label{plot_LFIR_Lradio_newfit}
\end{figure}
\begin{figure*}
  \includegraphics[width=\textwidth]{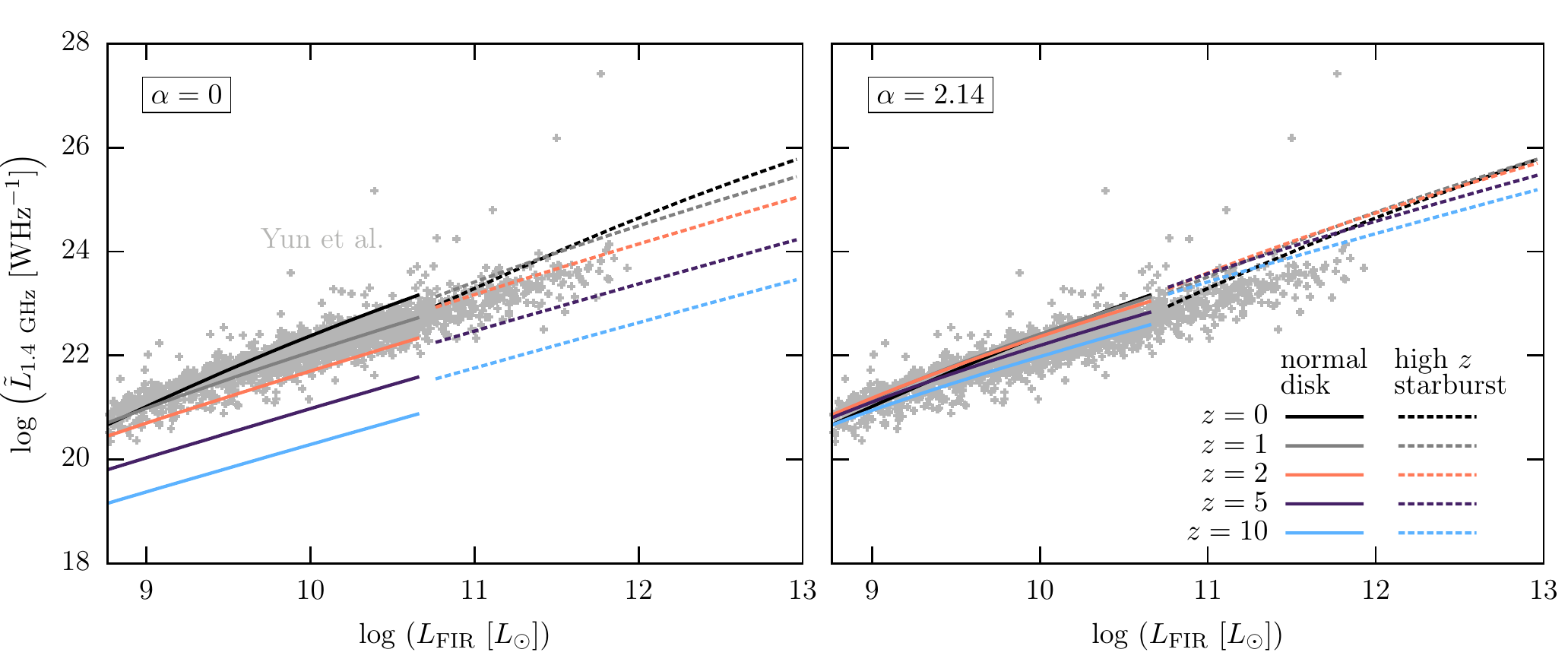}
  \caption{The radio luminosity density $L_{1.4~\mathrm{GHz}}$ as a function of the FIR luminosity $L_\mathrm{FIR}$ for different redshifts. We employ a model based on the Milky Way for $6\times10^{7}~L_\odot \lesssim L_\mathrm{FIR} \lesssim 6\times10^{10}~L_\odot$ and a model based on a high-$z$ starburst for $L_\mathrm{FIR} \gtrsim 6\times10^{10}~L_\odot$. For reference we overplot the data given in \citet{YunEtAl2001} which correspond to a redshift of 0. In the left panel we employ a gas density evolution with $\alpha=0$, while $\alpha=2.14$ in the right panel.}
\label{plot_LFIR_Lradio}
\end{figure*}
We use this relation to determine $L_\mathrm{FIR}$ from the observational data reported in \citet{YunEtAl2001}. \\
The FIR luminosity correlates with the radio luminosity in the local Universe \citep{YunEtAl2001}. As the origin of this correlation is a strong coupling between star formation and magnetic fields, the responsible contribution to the radio emission is the non-thermal synchrotron radiation. We calculate the latter using equation (\ref{eq_Lnu}). Observational studies often use a radio luminosity density at a given frequency instead of a luminosity integrated over a frequency range. We calculate the radio luminosity density $\tilde{L}_\mathrm{1.4~GHz} = L_\nu |_{\nu=1.4~\mathrm{GHz}}$ at a frequency of $\nu = 1.4$ GHz based on the spectral synchrotron luminosity derived in equation (\ref{eq_Lnu}). The radio luminosity at 1.4 GHz is calculated as $L_\mathrm{1.4~GHz} = 1.4~\mathrm{GHz}~\tilde{L}_\mathrm{1.4~GHz}$.\\
A qualitative comparison between our model and the observed data from \citet{YunEtAl2001} at $z=0$ is shown in figure \ref{plot_LFIR_Lradio_newfit}. We also show the results for different fit functions, linear ones and power-law fits. The fitting functions are all given in the figure. The prediction from our theoretical galaxy model is over plotted with the orange piecewise line. We emphasize once more that we did not perform any fit to the data with our model. All the parameters are chosen on the basis of information from our two fiducial galaxy models: the models of the Milky Way and the high-$z$ starburst galaxy. With the exception of very low $L_\mathrm{FIR}$, i.e.~very low star formation rates, our model reproduces the data at $z=0$ remarkably well. The discrepancy at low $\dot{M}_\star$ is a result of our assumption of relatively strong winds and the corresponding losses of cosmic ray electrons. The wind velocity is discussed in section \ref{sec_Model} and is based on the Milky Way with a weak dependence $\dot{M}_\star$ given in (\ref{eq_vwind}). For galaxies with low $\dot{M}_\star$ our estimate of the wind losses is probably too high, leading to a smaller number of cosmic rays and thus to less synchrotron emission. \\ 
In order to quantify the differences between the fits and the data and between the data and our model, we perform a goodness-of-fit test. For this we calculate the $R^2$ value. Assume that a model predicts an expected value $m_i$ for every data point $d_i$ with $i=1,...,n$ from a set of $n$ data points. Then the $R^2$ value is defined as
\begin{equation}
  R^2 = 1 - \frac{\sigma_\mathrm{res}}{\sigma_\mathrm{tot}}
\label{eq_R2}
\end{equation}
with the residual sum of squares $\sigma_\mathrm{res} \equiv \sum_i (m_i - \overline{d_i})^2$ and the total sum of squares $\sigma_\mathrm{tot} \equiv \sum_i (d_i - \overline{d_i})^2$. The mean value of the data is defined as $\overline{d_i} \equiv \sum_i d_i/n$. A value of $R^2 = 1$ would correspond to a model that fits the data perfectly, while a model with $R^2 = 0$ does not fit the data at all. The adjusted $R^2$ value $R^2_\mathrm{adj}$ takes into account the sample size $N$ and the number of variables in the model $p$. It is calculated as
\begin{equation}
  R^2_\mathrm{adj} = R^2  - (1 - R^2) \frac{p}{N-p-1}.
\label{eq_R2adj}
\end{equation}
We give the $R^2_\mathrm{adj}$ values for all the fits to the the FIR-radio correlation data in figure \ref{plot_LFIR_Lradio}. Our theoretical galaxy model has a value of $R^2_\mathrm{adj} = 0.759$ and is thus worse than the linear fit functions and the power-law fit. The smaller $R^2_\mathrm{adj}$ value compared to the linear fits can be traced back to the slope of our curve which does not match the data perfectly. Still as the $R^2_\mathrm{adj}$ for our model is high, we can consider it to reproduce the local observations well. We stress again that the orange curve in figure \ref{plot_LFIR_Lradio} is not based on a fit and that we cannot expect a perfect agreement with the data.

\subsection{At high redshift}
\label{sec_FIR-radio_highz}
While the FIR-radio correlation is well tested in the local Universe, its evolution with redshift is less clear. For example, \citet{MurphyEtAl2009a} cannot confirm any evolution of the correlation up to $z\approx2.6$ for their galaxy sample. They do however report that the median ratio of IR to radio flux in their submillimeter galaxies is lower than in local star forming galaxies by a factor of roughly 3. A physical model for the FIR-radio correlation as a function of redshift is proposed in \citet{Murphy2009}. In this work the synchrotron emission is calculated for fixed galactic magnetic fields. Unless the magnetic fields in high-$z$ star-forming galaxies are extremely strong, the author suggests that the ratio of infrared to radio luminosity increases with redshift. An additional model has been presented by \citet{LackiEtAl2010b}. With their one-zone galaxy model that includes a detailed cosmic ray description, they also predict an evolution of the FIR-radio correlation with redshift which is most significant for galaxies with low star formation rates. In this section we use our theoretical model for galaxies, their cosmic rays, and their dynamo-produced magnetic fields to predict this correlation at very high redshift. \\
We employ the model that was derived in section \ref{sec_Model} to calculate the radio luminosity density as a function of the FIR luminosity at different redshifts. Our estimate for $L_\mathrm{FIR}$ is given in equation (\ref{eq_LFIR}). We model the FIR luminosity only as a function of the star formation rate. Any possible redshift evolution of the dust is not taken into account in this work. Note that the transition from the Milky Way type regime to the high-$z$ starburst regime occurs at $\dot{M}_\star = 10~M_\odot\mathrm{yr}^{-1}$ which corresponds to $L_\mathrm{FIR} \approx 6\times10^{10}~L_\odot$. The resulting $z$-dependent correlation is shown in figure \ref{plot_LFIR_Lradio}. With increasing redshift $L_{1.4~\mathrm{GHz}}$ clearly decreases for a given $L_\mathrm{FIR}$. This originates from the decreasing energy loss timescale of cosmic ray electrons (see figure \ref{plot_timescales_SFR2}). The results are fewer cosmic rays thus less synchrotron emission. Moreover, our model predicts the FIR-radio correlation to flatten slightly with increasing redshift. In figure \ref{plot_LFIR_Lradio} the difference between a strong and a weak evolution of the gas density (\ref{eq_n}) with redshift is shown. In the left panel $\alpha=0$ which corresponds to $n\propto(1+z)^{5.14}$. Here at a redshift of 2 (see the orange line in figure \ref{plot_LFIR_Lradio}) the FIR-radio correlation lies below the observational data in the local Universe. For $\alpha=2.14$ which implies $n\propto(1+z)^{3}$ the evolution of the FIR-radio correlation is not significant and the model curves below $z\approx5$ can hardly be distinguished. \\
A breakdown of the FIR-radio correlation at high redshift has been already proposed by \citet{SchleicherBeck2013}.
\begin{figure*}
  \includegraphics[width=\textwidth]{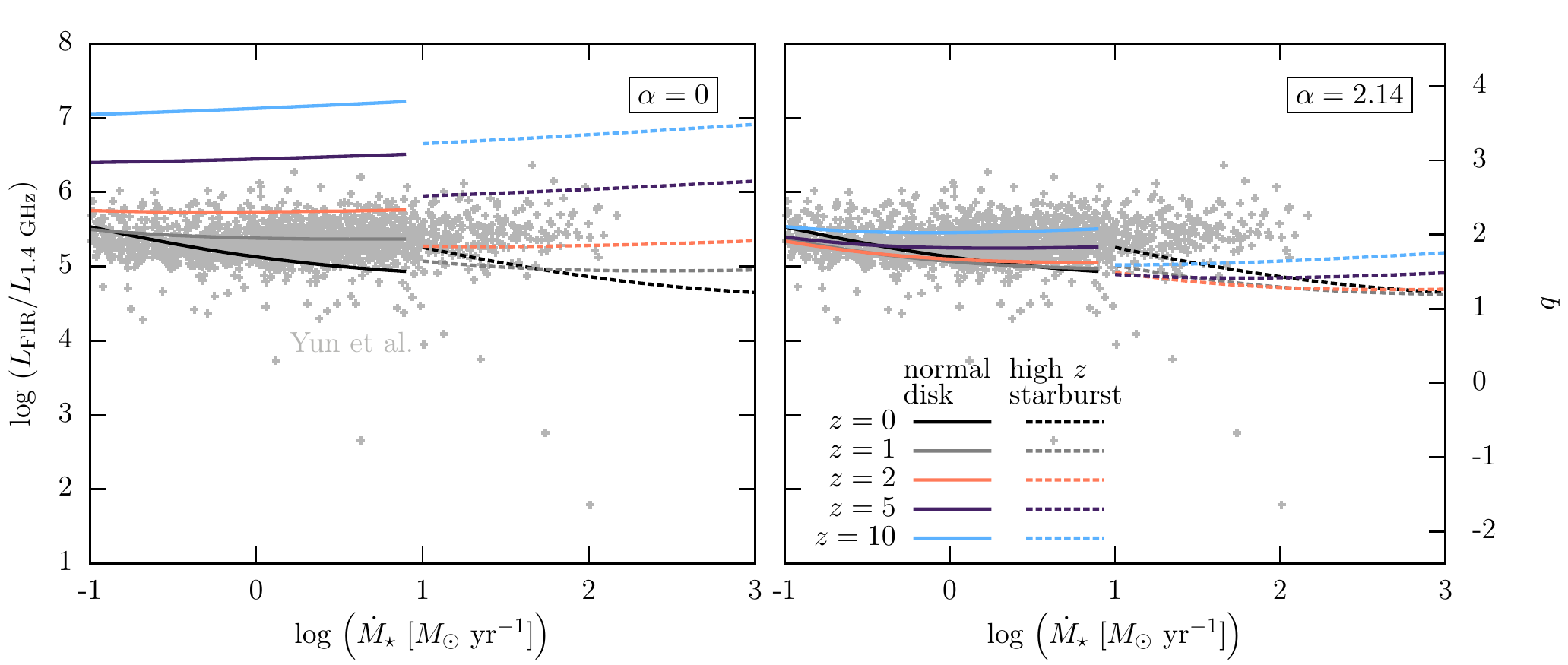}
  \caption{The ratio of the FIR luminosity $L_\mathrm{FIR}$ over radio luminosity $L_{1.4~\mathrm{GHz}}$  as a function of star formation rate $\dot{M}_\star$. We employ our model based on a Milky Way like disk for $\dot{M}_\star \le 10~M_\odot\mathrm{yr}^{-1}$ and the one based on the high-$z$ starburst for larger $\dot{M}_\star$. Different redshifts are indicated by different colors. As in figure \ref{plot_LFIR_Lradio} the left panel shows $\alpha=0$ and the right one $\alpha=2.14$. }
\label{plot_LFIRLradio_SFR}
\end{figure*}
This study is also based on the assumption that galactic magnetic fields are produced by a small-scale turbulent dynamo and thus depend on the star formation rate. \citet{SchleicherBeck2013} estimate the typical timescales for energy losses of cosmic ray electrons. The loss timescales they assume at a redshift of zero are similar to the ones we employ. The most significant difference is the timescale for bremsstrahlung which is smaller by a factor of three in our case. In addition we include losses by galactic winds that are in our model dominant for a large range of $\dot{M}_\star$ at $z=0$ and low cosmic ray energies (see figure \ref{plot_timescales_SFR2}). Above a critical redshift that is defined by the time where $\tau_\mathrm{IC}$ becomes smaller then $\tau_\mathrm{synch}$, \citet{SchleicherBeck2013} predict a breakdown of the FIR-radio correlation. In our work we employ a strong evolution of the number density with redshift which leads to a fast decrease of $\tau_\mathrm{brems}$ and $\tau_\mathrm{ion}$ with redshift. For low cosmic ray energies the latter become the dominant loss channels already at moderate $z$. Even though synchrotron losses are not dominating there is still radio emission. We extend the approach by \citet{SchleicherBeck2013} and calculate the expected synchrotron emission based on the steady state number density of cosmic ray electrons. In agreement with their prediction we find that $\tau_\mathrm{IC}$ decreases rapidly with redshift and that synchrotron emission is not the dominate loss channel in young galaxies. Instead of a breakdown of the FIR-radio correlation we find an evolution in cosmic time where non-thermal radio emission significantly decreases with redshift. Due to our focus on starburst galaxies with strong, i.e.~efficiently amplified magnetic fields, the inverse Compton scattering is less relevant than other loss mechanisms, which may change when probing more typical galaxies. Such variations in the overall efficiency in the production of non-thermal synchrotron are not only relevant as probes of the magnetic fields, but also when employing radio emission as a tracer of star formation. \\ 
We note, that our model results in an evolution of the radio luminosity over the whole range of FIR luminosities studied here. \citet{LackiEtAl2010b} on the other hand propose an evolution of the FIR-radio correlation that is only significant at low star formation rates. The origin of this discrepancy lies probably in the different models of the density. While there is a scaling with the star formation rate in both studies (see our equation \ref{eq_n}), \citet{LackiEtAl2010b} do not include a scaling of $n$ with redshift. With our scaling of $n\propto (1+z)^{3-\alpha}$ that is motivated from the evolution of the galactic volume (\ref{eq_Vgal}) the timescales of ionization and bremstrahlung decrease strongly with redshift (see figure \ref{plot_timescales_SFR2}). We are not studying the regime where inverse Compton scattering is the dominant energy loss of cosmic ray electrons. Losses would only be dominated by inverse Compton scattering if the magnetic fields were weak at high redshift. Assuming the presence of a small-scale dynamo, we expect that magnetic fields in young galaxies are very strong. \\
Similar conclusions can be drawn from figure \ref{plot_LFIRLradio_SFR} where we plot the ratio of FIR over radio luminosity as a function of the star formation rate. In case of a linear FIR-radio correlation the curves should be horizontal. Our model curves clearly deviate from a linear correlation. In fact, the ratio decreases with $\dot{M}_\star$ for low redshifts up to very high $\dot{M}_\star$. However, at higher redshifts $L_\mathrm{FIR}/(L_{1.4~\mathrm{GHz}})$ grows with the star formation rate. We illustrate this trend also in figure \ref{plot_LFIRLradio_z}, where we show the same ratio as a function of $z$ for our two fiducial galaxy models. In both cases the fraction $L_\mathrm{FIR}/(L_{1.4~\mathrm{GHz}})$ has increased by an order of magnitude already at a redshift of $z=2$ for $\alpha=0$. If the density increases only as $(1+z)^3$, the evolution of $L_\mathrm{FIR}/(L_{1.4~\mathrm{GHz}})$ is less significant. In fact, the difference between the ratio in local Universe and at $z=10$ is less than an order of magnitude. \\
For better comparison with the literature we indicate the so-called $q$ parameter on the right hand axes of figure \ref{plot_LFIRLradio_SFR}. This quantity is defined as \citep{HelouEtAl1985}
\begin{equation}
  q = \mathrm{log}\left(\frac{S_\mathrm{FIR}}{3.75\times10^{12}~\mathrm{Hz}}\right) - \mathrm{log}\left(\tilde{S}_{1.4~\mathrm{GHz}}\right),
\label{eq_q}
\end{equation}
where $S_\mathrm{FIR}$ is the FIR flux, $\tilde{S}_{1.4~\mathrm{GHz}}$ the 1.4 GHz flux density, and $3.75\times10^{12}~\mathrm{Hz}$ the frequency at 80 $\mu$m. In terms of luminosities we find 
\begin{equation}
  q = \mathrm{log}\left(\frac{L_\mathrm{FIR}}{L_\mathrm{1.4~\mathrm{GHz}}}\right) - 3.43.
\label{eq_q2}
\end{equation}
Our model results in a $q$ parameter of roughly 2 at a redshift of $z=0$. We note that \citet{YunEtAl2001} report $q\approx2.3$, thus a value that is higher then indicated in our figure. This can be caused by their calculation of the FIR luminosity which includes not only the 60 $\mu$m flux that we use here, but additionally the flux at 100 $\mu$m. In addition, \citet{YunEtAl2001} did not substract the contribution from free-free emission to the radio spectrum which could lead to an overestimate of the non-thermal emission. \\
A main result of our analysis is an increase of the $q$ parameter with redshift for galaxy sample with similar star formation rates. The evolution of $q$ is also predicted by the theoretical model in \citet{LackiEtAl2010b}. So far there is no agreement on a possible evolution of this parameter from the observational side. While some works report no evolution of the FIR-radio correlation up to moderate redshifts \citep{MurphyEtAl2009a,BargerEtAl2012} others even find a decrease of $q$ \citep{IvisonEtAl2010b,BourneEtAl2011,BasuEtAl2015}. These observations, however, typically cover galaxies with different star formation rates while in figure \ref{plot_LFIRLradio_z} galaxies with fixed $\dot{M}_\star$ are considered. At high redshift the mean $\dot{M}_\star$ is higher \citep{MadauEtAl1998} plus observations are easier for brighter galaxies. In addition the presence of active galactic nuclei will bias observational results towards a lower $q$.



\section{Conclusions}
\label{sec_Conclusions}
Based on the assumption that galaxies have had strong magnetic fields throughout their evolution we calculate their synchrotron emission. We argue that the origin of these strong fields is rapid amplification of seed fields via a turbulent small-scale dynamo. With the energy input for the dynamo being turbulence there is a direct coupling to the star formation activity in galaxies. We thus can expect the magnetic field strength of a galaxy to scale with its star formation rate. This relation is quantified in equation (\ref{eq_B}). 
\begin{figure}[t]
  \includegraphics[width=0.5\textwidth]{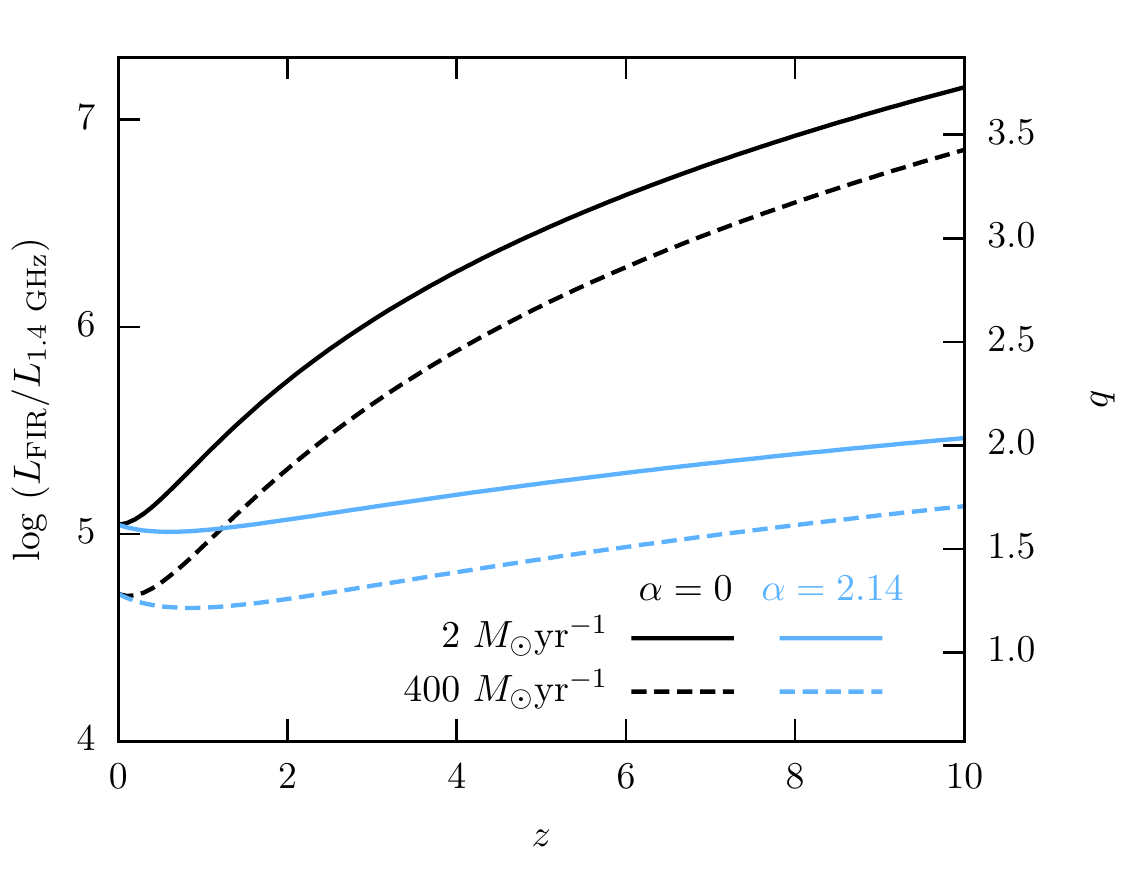}
  \caption{The ratio of the FIR luminosity $L_\mathrm{FIR}$ over radio luminosity $L_{1.4~\mathrm{GHz}}$ as a function of $z$. We present the result for our two fiducial galaxy models: a model of the Milky Way with a star formation rate of $2~M_\odot\mathrm{yr}^{-1}$ (solid lines) and the high-$z$ starburst galaxy with $\dot{M}_\star = 400~M_\odot\mathrm{yr}^{-1}$ (dashed lines). The black lines correspond the $\alpha=0$ in the gas density evolution, while the blue lines refer to $\alpha=2.14$.}
\label{plot_LFIRLradio_z}
\end{figure}
A tracer of magnetic fields is synchrotron emission from cosmic rays. The origin of these highly energetic particles are shock fronts in expanding supernova shells which again are coupled to the star formation rate (see equation \ref{eq_NSN}). \\
In this study we construct a semi-analytical galaxy model which covers the evolution of the volume, density, magnetic field, and cosmic rays of a galaxy. We distinguish two different regimes of star formation rate. For $\dot{M}_\star < 10~M_\odot\mathrm{yr}^{-1}$ our model is based on the Milky Way, while we employ SMM J2135-0102, a starburst galaxy at $z=2.3$, as a template for $\dot{M}_\star > 10~M_\odot\mathrm{yr}^{-1}$. The details are discussed in section \ref{sec_Model}. Our galaxy model includes several assumptions. First of all we employ two exemplary galaxies to describe a huge range of star formation rates. These galaxies are homogeneous disks with radii and scale heights that depend only on redshift. In nature there is of course a variety of different galaxy morphologies. Further, the two regimes of star formation rates are modeled based on the interstellar radiation fields of the fiducial galaxies. Although we include a scaling with the star formation rate density, real interstellar radiation fields might be more complex. The cosmic ray model we use neglects diffusion which is caused by the assumption of homogeneity. Moreover we focus solely on the injection of cosmic rays by supernovae. The latter are also our only source of turbulence. Our one-zone galaxy model can thus only predict trends for the synchrotron emission and the evolution of the FIR-radio correlation. The quantitative uncertainties may therefore be as large as one order of magnitude.\\
The main results of this study are summarized in the following:
\begin{itemize}
\item{The number of cosmic ray electrons in steady state depends on various cooling mechanisms. The cooling timescale is a function of the star formation rate as well as of the redshift. At high redshifts cosmic ray losses are dominated by ionization and bremsstrahlung (see figure \ref{plot_timescales_SFR2}). The latter depend on the particle density which is directly coupled to the star formation rate density (see equation \ref{eq_n}).}
\item{With our cosmic ray model and a scaling of the magnetic field strength with the star formation rate that remains similar at high redshift we calculate the galactic synchrotron emission. The results for a Milky Way like galaxy and a typical high-$z$ starburst are shown in figure \ref{plot_Snu_nu}. We compare here the observed synchrotron flux with conservative sensitivity values of current radio telescopes. We find that detection of non-thermal radio emission is possible up to $z\approx2$. Especially at high redshift free-free absorption at low and free-free emission at high $\nu$ makes multi-frequency observations necessary.}
\item{We calculate the radio luminosity density at $\nu=1.4$ GHz and estimate the corresponding FIR luminosity using the relation to the star formation rate from \citet{Kennicutt1998}. The resulting FIR-radio correlation is presented in figure \ref{plot_LFIR_Lradio_newfit}. Here we compare our theoretical prediction with the observational data of local galaxies from \citet{YunEtAl2001}. Without adjusting the parameters of our model we find good agreement with the observational data at $z=0$.}
\item{We study the evolution of the FIR-radio correlation with redshift in figures \ref{plot_LFIR_Lradio}, \ref{plot_LFIRLradio_SFR}, and \ref{plot_LFIRLradio_z}. Due to the increasing energy losses of cosmic ray electrons the radio luminosity decreases with redshift. This leads to a smaller FIR-radio correlation for younger galaxies. Moreover, the slope of the correlation becomes smaller with increasing $z$. How much the correlation changes depends strongly on the scaling of the gas density $n\propto(1+z)^{5.14-\alpha}$. For $\alpha=0$ our model predicts a significant evolution of the FIR-radio correlation with redshift over the entire range of star formation rates that is already visible at $z\approx2$. If the $z$-dependence of the density is weaker, i.e.~$\alpha=2.14$, a modification of the correlation could only be detected at $z\gtrsim5$.}
\end{itemize}
Our theoretical galaxy model should be testable in future radio observations. A current survey with the JVLA is CHILES \textit{con Pol}\footnote{http://www.chilesconpol.com/} which will provide the deepest window into the radio sky up to date. The survey covers the COSMOS\footnote{http://cosmos.astro.caltech.edu/} field and thus deep observations in various wavelengths are already available. CHILES \textit{con Pol} will be able to test the FIR-radio correlation up to cosmological redshifts. With a high angular resolution and a enormous sensitivity the SKA as well as its pathfinders like LOFAR and ASKAP will be powerful telescopes to observe galactic radio emission in the early Universe. The expected sensitivity of the SKA\textit{-mid} is roughly 0.3 $\mu$Jy for an exposure time of 8 hours. The synchrotron emission from typical high-$z$ starburst galaxies, like our fiducial galaxy SMM J2135-0102, can be detected up to $z\approx2$ according to our theoretical model for the evolution of cosmic rays and galactic magnetic fields. For longer integration times the flux sensitivity decreases and radiation from more distant sources can be observed. For instance an 100 hour integration would push the sensitivity to 0.07 $\mu$Jy and magnetic fields at a redshift of 5 and above could be studied. \\
The combination of new radio observations and theoretical models like the one presented in this work will be important for better understanding of galaxies. In particular, we will gain more insights in the evolution of galactic magnetic fields, cosmic rays, and star formation which are closely coupled. 


\acknowledgements{We thank Christopher Hales for discussions about current and future radio observations of the FIR-radio correlation. We are also grateful to the anonymous referee for his thoughtful suggestions. This work has been financially supported by {\em Nordita} which is funded by the Nordic Council of Ministers, the Swedish Research Council, and the two host universities, the {\em Royal Institute of Technology} (KTH) and {\em Stockholm University}. DRGS thanks for funding through Fondecyt regular (project code 1161247) and through the ``Concurso Proyectos Internacionales de Investigaci\'on, Convocatoria 2015'' (project code PII20150171). We further thank for funding through the {\em Deutsche Forschungsgemeinschaft} (DFG) in the {\em Schwer\-punkt\-programm} SPP 1573 ``Physics of the Interstellar Medium'' under grants KL 1358/14-1, SCHL 1964/1-1, SCHL 1964/1-2, and BO 4113/1-2. In addition we thank the DFG for support via the SFB 881 ``The Milky Way System'' in the sub-projects B1 and B2. In addition we acknowledge financial support by the {\em European Research Council} under the {\em European Community's Seventh Framework Programme} (FP7/2007-2013) via the ERC {\em Advanced Grant} STARLIGHT (project number 339177).}


\begin{thebibliography}{75}%
\makeatletter
\providecommand \@ifxundefined [1]{%
 \@ifx{#1\undefined}
}%
\providecommand \@ifnum [1]{%
 \ifnum #1\expandafter \@firstoftwo
 \else \expandafter \@secondoftwo
 \fi
}%
\providecommand \@ifx [1]{%
 \ifx #1\expandafter \@firstoftwo
 \else \expandafter \@secondoftwo
 \fi
}%
\providecommand \natexlab [1]{#1}%
\providecommand \enquote  [1]{``#1''}%
\providecommand \bibnamefont  [1]{#1}%
\providecommand \bibfnamefont [1]{#1}%
\providecommand \citenamefont [1]{#1}%
\providecommand \href@noop [0]{\@secondoftwo}%
\providecommand \href [0]{\begingroup \@sanitize@url \@href}%
\providecommand \@href[1]{\@@startlink{#1}\@@href}%
\providecommand \@@href[1]{\endgroup#1\@@endlink}%
\providecommand \@sanitize@url [0]{\catcode `\\12\catcode `\$12\catcode
  `\&12\catcode `\#12\catcode `\^12\catcode `\_12\catcode `\%12\relax}%
\providecommand \@@startlink[1]{}%
\providecommand \@@endlink[0]{}%
\providecommand \url  [0]{\begingroup\@sanitize@url \@url }%
\providecommand \@url [1]{\endgroup\@href {#1}{\urlprefix }}%
\providecommand \urlprefix  [0]{URL }%
\providecommand \Eprint [0]{\href }%
\providecommand \doibase [0]{http://dx.doi.org/}%
\providecommand \selectlanguage [0]{\@gobble}%
\providecommand \bibinfo  [0]{\@secondoftwo}%
\providecommand \bibfield  [0]{\@secondoftwo}%
\providecommand \translation [1]{[#1]}%
\providecommand \BibitemOpen [0]{}%
\providecommand \bibitemStop [0]{}%
\providecommand \bibitemNoStop [0]{.\EOS\space}%
\providecommand \EOS [0]{\spacefactor3000\relax}%
\providecommand \BibitemShut  [1]{\csname bibitem#1\endcsname}%
\let\auto@bib@innerbib\@empty
\bibitem [{\citenamefont {{Beck}}(2011)}]{Beck2011}%
  \BibitemOpen
  \bibfield  {author} {\bibinfo {author} {\bibfnamefont {R.}~\bibnamefont
  {{Beck}}},\ }\href {\doibase 10.1007/s11214-011-9782-z} {\bibfield  {journal}
  {\bibinfo  {journal} {\ssr}\ ,\ \bibinfo {pages} {135}} (\bibinfo {year}
  {2011})}\BibitemShut {NoStop}%
\bibitem [{\citenamefont {Turner}\ and\ \citenamefont
  {Widrow}(1988)}]{TurnerWidrow1988}%
  \BibitemOpen
  \bibfield  {author} {\bibinfo {author} {\bibfnamefont {M.~S.}\ \bibnamefont
  {Turner}}\ and\ \bibinfo {author} {\bibfnamefont {L.~M.}\ \bibnamefont
  {Widrow}},\ }\href {\doibase 10.1103/PhysRevD.37.2743} {\bibfield  {journal}
  {\bibinfo  {journal} {Phys.~Rev.~D}\ }\textbf {\bibinfo {volume} {37}},\
  \bibinfo {pages} {2743} (\bibinfo {year} {1988})}\BibitemShut {NoStop}%
\bibitem [{\citenamefont {{Ratra}}(1992)}]{Ratra1992}%
  \BibitemOpen
  \bibfield  {author} {\bibinfo {author} {\bibfnamefont {B.}~\bibnamefont
  {{Ratra}}},\ }\href {\doibase 10.1086/186384} {\bibfield  {journal} {\bibinfo
   {journal} {\apjl}\ }\textbf {\bibinfo {volume} {391}},\ \bibinfo {pages}
  {L1} (\bibinfo {year} {1992})}\BibitemShut {NoStop}%
\bibitem [{\citenamefont {{Quashnock}}\ \emph {et~al.}(1989)\citenamefont
  {{Quashnock}}, \citenamefont {{Loeb}},\ and\ \citenamefont
  {{Spergel}}}]{QuashnockEtAl1989}%
  \BibitemOpen
  \bibfield  {author} {\bibinfo {author} {\bibfnamefont {J.~M.}\ \bibnamefont
  {{Quashnock}}}, \bibinfo {author} {\bibfnamefont {A.}~\bibnamefont {{Loeb}}},
  \ and\ \bibinfo {author} {\bibfnamefont {D.~N.}\ \bibnamefont {{Spergel}}},\
  }\href {\doibase 10.1086/185528} {\bibfield  {journal} {\bibinfo  {journal}
  {\apjl}\ }\textbf {\bibinfo {volume} {344}},\ \bibinfo {pages} {L49}
  (\bibinfo {year} {1989})}\BibitemShut {NoStop}%
\bibitem [{\citenamefont {{Sigl}}\ \emph {et~al.}(1997)\citenamefont {{Sigl}},
  \citenamefont {{Olinto}},\ and\ \citenamefont {{Jedamzik}}}]{SiglEtAl1997}%
  \BibitemOpen
  \bibfield  {author} {\bibinfo {author} {\bibfnamefont {G.}~\bibnamefont
  {{Sigl}}}, \bibinfo {author} {\bibfnamefont {A.~V.}\ \bibnamefont
  {{Olinto}}}, \ and\ \bibinfo {author} {\bibfnamefont {K.}~\bibnamefont
  {{Jedamzik}}},\ }\href {\doibase 10.1103/PhysRevD.55.4582} {\bibfield
  {journal} {\bibinfo  {journal} {\prd}\ }\textbf {\bibinfo {volume} {55}},\
  \bibinfo {pages} {4582} (\bibinfo {year} {1997})}\BibitemShut {NoStop}%
\bibitem [{\citenamefont {{Xu}}\ \emph {et~al.}(2008)\citenamefont {{Xu}},
  \citenamefont {{O'Shea}}, \citenamefont {{Collins}}, \citenamefont
  {{Norman}}, \citenamefont {{Li}},\ and\ \citenamefont {{Li}}}]{XuEtAl2008}%
  \BibitemOpen
  \bibfield  {author} {\bibinfo {author} {\bibfnamefont {H.}~\bibnamefont
  {{Xu}}}, \bibinfo {author} {\bibfnamefont {B.~W.}\ \bibnamefont {{O'Shea}}},
  \bibinfo {author} {\bibfnamefont {D.~C.}\ \bibnamefont {{Collins}}}, \bibinfo
  {author} {\bibfnamefont {M.~L.}\ \bibnamefont {{Norman}}}, \bibinfo {author}
  {\bibfnamefont {H.}~\bibnamefont {{Li}}}, \ and\ \bibinfo {author}
  {\bibfnamefont {S.}~\bibnamefont {{Li}}},\ }\href {\doibase 10.1086/595617}
  {\bibfield  {journal} {\bibinfo  {journal} {\apjl}\ }\textbf {\bibinfo
  {volume} {688}},\ \bibinfo {pages} {L57} (\bibinfo {year}
  {2008})}\BibitemShut {NoStop}%
\bibitem [{\citenamefont {{Schlickeiser}}(2012)}]{Schlickeiser2012}%
  \BibitemOpen
  \bibfield  {author} {\bibinfo {author} {\bibfnamefont {R.}~\bibnamefont
  {{Schlickeiser}}},\ }\href {\doibase 10.1103/PhysRevLett.109.261101}
  {\bibfield  {journal} {\bibinfo  {journal} {\prl}\ }\textbf {\bibinfo
  {volume} {109}},\ \bibinfo {eid} {261101} (\bibinfo {year}
  {2012})}\BibitemShut {NoStop}%
\bibitem [{\citenamefont {{Beck}}(2001)}]{Beck2001}%
  \BibitemOpen
  \bibfield  {author} {\bibinfo {author} {\bibfnamefont {R.}~\bibnamefont
  {{Beck}}},\ }\href@noop {} {\bibfield  {journal} {\bibinfo  {journal} {\ssr}\
  }\textbf {\bibinfo {volume} {99}},\ \bibinfo {pages} {243} (\bibinfo {year}
  {2001})}\BibitemShut {NoStop}%
\bibitem [{\citenamefont {{Brandenburg}}\ and\ \citenamefont
  {{Subramanian}}(2005)}]{BrandenburgSubramanian2005}%
  \BibitemOpen
  \bibfield  {author} {\bibinfo {author} {\bibfnamefont {A.}~\bibnamefont
  {{Brandenburg}}}\ and\ \bibinfo {author} {\bibfnamefont {K.}~\bibnamefont
  {{Subramanian}}},\ }\href {\doibase 10.1016/j.physrep.2005.06.005} {\bibfield
   {journal} {\bibinfo  {journal} {\physrep}\ }\textbf {\bibinfo {volume}
  {417}},\ \bibinfo {pages} {1} (\bibinfo {year} {2005})}\BibitemShut {NoStop}%
\bibitem [{\citenamefont {Federrath}\ \emph {et~al.}(2011)\citenamefont
  {Federrath}, \citenamefont {Chabrier}, \citenamefont {Schober}, \citenamefont
  {Banerjee}, \citenamefont {Klessen},\ and\ \citenamefont
  {Schleicher}}]{FederrathEtAl2011b}%
  \BibitemOpen
  \bibfield  {author} {\bibinfo {author} {\bibfnamefont {C.}~\bibnamefont
  {Federrath}}, \bibinfo {author} {\bibfnamefont {G.}~\bibnamefont {Chabrier}},
  \bibinfo {author} {\bibfnamefont {J.}~\bibnamefont {Schober}}, \bibinfo
  {author} {\bibfnamefont {R.}~\bibnamefont {Banerjee}}, \bibinfo {author}
  {\bibfnamefont {R.~S.}\ \bibnamefont {Klessen}}, \ and\ \bibinfo {author}
  {\bibfnamefont {D.~R.~G.}\ \bibnamefont {Schleicher}},\ }\href {\doibase
  10.1103/PhysRevLett.107.114504} {\bibfield  {journal} {\bibinfo  {journal}
  {\prl}\ }\textbf {\bibinfo {volume} {107}},\ \bibinfo {pages} {114504}
  (\bibinfo {year} {2011})}\BibitemShut {NoStop}%
\bibitem [{\citenamefont {{Schober}}\ \emph
  {et~al.}(2012{\natexlab{a}})\citenamefont {{Schober}}, \citenamefont
  {{Schleicher}}, \citenamefont {{Federrath}}, \citenamefont {{Klessen}},\ and\
  \citenamefont {{Banerjee}}}]{SchoberEtAl2012.1}%
  \BibitemOpen
  \bibfield  {author} {\bibinfo {author} {\bibfnamefont {J.}~\bibnamefont
  {{Schober}}}, \bibinfo {author} {\bibfnamefont {D.}~\bibnamefont
  {{Schleicher}}}, \bibinfo {author} {\bibfnamefont {C.}~\bibnamefont
  {{Federrath}}}, \bibinfo {author} {\bibfnamefont {R.}~\bibnamefont
  {{Klessen}}}, \ and\ \bibinfo {author} {\bibfnamefont {R.}~\bibnamefont
  {{Banerjee}}},\ }\href {\doibase 10.1103/PhysRevE.85.026303} {\bibfield
  {journal} {\bibinfo  {journal} {\pre}\ }\textbf {\bibinfo {volume} {85}},\
  \bibinfo {eid} {026303} (\bibinfo {year} {2012}{\natexlab{a}})}\BibitemShut
  {NoStop}%
\bibitem [{\citenamefont {{Schober}}\ \emph
  {et~al.}(2012{\natexlab{b}})\citenamefont {{Schober}}, \citenamefont
  {{Schleicher}}, \citenamefont {{Bovino}},\ and\ \citenamefont
  {{Klessen}}}]{SchoberEtAl2012.3}%
  \BibitemOpen
  \bibfield  {author} {\bibinfo {author} {\bibfnamefont {J.}~\bibnamefont
  {{Schober}}}, \bibinfo {author} {\bibfnamefont {D.}~\bibnamefont
  {{Schleicher}}}, \bibinfo {author} {\bibfnamefont {S.}~\bibnamefont
  {{Bovino}}}, \ and\ \bibinfo {author} {\bibfnamefont {R.~S.}\ \bibnamefont
  {{Klessen}}},\ }\href {\doibase 10.1103/PhysRevE.86.066412} {\bibfield
  {journal} {\bibinfo  {journal} {\pre}\ }\textbf {\bibinfo {volume} {86}},\
  \bibinfo {eid} {066412} (\bibinfo {year} {2012}{\natexlab{b}})}\BibitemShut
  {NoStop}%
\bibitem [{\citenamefont {{Federrath}}\ \emph {et~al.}(2014)\citenamefont
  {{Federrath}}, \citenamefont {{Schober}}, \citenamefont {{Bovino}},\ and\
  \citenamefont {{Schleicher}}}]{FederrathEtAl2014b}%
  \BibitemOpen
  \bibfield  {author} {\bibinfo {author} {\bibfnamefont {C.}~\bibnamefont
  {{Federrath}}}, \bibinfo {author} {\bibfnamefont {J.}~\bibnamefont
  {{Schober}}}, \bibinfo {author} {\bibfnamefont {S.}~\bibnamefont {{Bovino}}},
  \ and\ \bibinfo {author} {\bibfnamefont {D.~R.~G.}\ \bibnamefont
  {{Schleicher}}},\ }\href {\doibase 10.1088/2041-8205/797/2/L19} {\bibfield
  {journal} {\bibinfo  {journal} {\apjl}\ }\textbf {\bibinfo {volume} {797}},\
  \bibinfo {eid} {L19} (\bibinfo {year} {2014})}\BibitemShut {NoStop}%
\bibitem [{\citenamefont {{Schober}}\ \emph {et~al.}(2013)\citenamefont
  {{Schober}}, \citenamefont {{Schleicher}},\ and\ \citenamefont
  {{Klessen}}}]{SchoberEtAl2013}%
  \BibitemOpen
  \bibfield  {author} {\bibinfo {author} {\bibfnamefont {J.}~\bibnamefont
  {{Schober}}}, \bibinfo {author} {\bibfnamefont {D.~R.~G.}\ \bibnamefont
  {{Schleicher}}}, \ and\ \bibinfo {author} {\bibfnamefont {R.~S.}\
  \bibnamefont {{Klessen}}},\ }\href {\doibase 10.1051/0004-6361/201322185}
  {\bibfield  {journal} {\bibinfo  {journal} {\aap}\ }\textbf {\bibinfo
  {volume} {560}},\ \bibinfo {eid} {A87} (\bibinfo {year} {2013})}\BibitemShut
  {NoStop}%
\bibitem [{\citenamefont {{Latif}}\ \emph {et~al.}(2013)\citenamefont
  {{Latif}}, \citenamefont {{Schleicher}}, \citenamefont {{Schmidt}},\ and\
  \citenamefont {{Niemeyer}}}]{LatifEtAl2013}%
  \BibitemOpen
  \bibfield  {author} {\bibinfo {author} {\bibfnamefont {M.~A.}\ \bibnamefont
  {{Latif}}}, \bibinfo {author} {\bibfnamefont {D.~R.~G.}\ \bibnamefont
  {{Schleicher}}}, \bibinfo {author} {\bibfnamefont {W.}~\bibnamefont
  {{Schmidt}}}, \ and\ \bibinfo {author} {\bibfnamefont {J.}~\bibnamefont
  {{Niemeyer}}},\ }\href {\doibase 10.1093/mnras/stt503} {\bibfield  {journal}
  {\bibinfo  {journal} {\mnras}\ } (\bibinfo {year} {2013}),\
  10.1093/mnras/stt503}\BibitemShut {NoStop}%
\bibitem [{\citenamefont {{Pakmor}}\ \emph {et~al.}(2014)\citenamefont
  {{Pakmor}}, \citenamefont {{Marinacci}},\ and\ \citenamefont
  {{Springel}}}]{PakmorMarinacciSpringel2014}%
  \BibitemOpen
  \bibfield  {author} {\bibinfo {author} {\bibfnamefont {R.}~\bibnamefont
  {{Pakmor}}}, \bibinfo {author} {\bibfnamefont {F.}~\bibnamefont
  {{Marinacci}}}, \ and\ \bibinfo {author} {\bibfnamefont {V.}~\bibnamefont
  {{Springel}}},\ }\href {\doibase 10.1088/2041-8205/783/1/L20} {\bibfield
  {journal} {\bibinfo  {journal} {\apjl}\ }\textbf {\bibinfo {volume} {783}},\
  \bibinfo {eid} {L20} (\bibinfo {year} {2014})}\BibitemShut {NoStop}%
\bibitem [{\citenamefont {{Beck}}\ and\ \citenamefont
  {{Krause}}(2005)}]{BeckKrause2005}%
  \BibitemOpen
  \bibfield  {author} {\bibinfo {author} {\bibfnamefont {R.}~\bibnamefont
  {{Beck}}}\ and\ \bibinfo {author} {\bibfnamefont {M.}~\bibnamefont
  {{Krause}}},\ }\href {\doibase 10.1002/asna.200510366} {\bibfield  {journal}
  {\bibinfo  {journal} {Astronomische Nachrichten}\ }\textbf {\bibinfo {volume}
  {326}},\ \bibinfo {pages} {414} (\bibinfo {year} {2005})}\BibitemShut
  {NoStop}%
\bibitem [{\citenamefont {{Yun}}\ \emph {et~al.}(2001)\citenamefont {{Yun}},
  \citenamefont {{Reddy}},\ and\ \citenamefont {{Condon}}}]{YunEtAl2001}%
  \BibitemOpen
  \bibfield  {author} {\bibinfo {author} {\bibfnamefont {M.~S.}\ \bibnamefont
  {{Yun}}}, \bibinfo {author} {\bibfnamefont {N.~A.}\ \bibnamefont {{Reddy}}},
  \ and\ \bibinfo {author} {\bibfnamefont {J.~J.}\ \bibnamefont {{Condon}}},\
  }\href {\doibase 10.1086/323145} {\bibfield  {journal} {\bibinfo  {journal}
  {\apj}\ }\textbf {\bibinfo {volume} {554}},\ \bibinfo {pages} {803} (\bibinfo
  {year} {2001})}\BibitemShut {NoStop}%
\bibitem [{\citenamefont {{Schleicher}}\ and\ \citenamefont
  {{Beck}}(2013)}]{SchleicherBeck2013}%
  \BibitemOpen
  \bibfield  {author} {\bibinfo {author} {\bibfnamefont {D.~R.~G.}\
  \bibnamefont {{Schleicher}}}\ and\ \bibinfo {author} {\bibfnamefont
  {R.}~\bibnamefont {{Beck}}},\ }\href {\doibase 10.1051/0004-6361/201321707}
  {\bibfield  {journal} {\bibinfo  {journal} {\aap}\ }\textbf {\bibinfo
  {volume} {556}},\ \bibinfo {eid} {A142} (\bibinfo {year} {2013})}\BibitemShut
  {NoStop}%
\bibitem [{\citenamefont {{Kennicutt}}(1998)}]{Kennicutt1998}%
  \BibitemOpen
  \bibfield  {author} {\bibinfo {author} {\bibfnamefont {R.~C.}\ \bibnamefont
  {{Kennicutt}}, \bibfnamefont {Jr.}},\ }\href {\doibase 10.1086/305588}
  {\bibfield  {journal} {\bibinfo  {journal} {\apj}\ }\textbf {\bibinfo
  {volume} {498}},\ \bibinfo {pages} {541} (\bibinfo {year}
  {1998})}\BibitemShut {NoStop}%
\bibitem [{\citenamefont {{Mac Low}}\ and\ \citenamefont
  {{Klessen}}(2004)}]{MacLowKlessen2004}%
  \BibitemOpen
  \bibfield  {author} {\bibinfo {author} {\bibfnamefont {M.-M.}\ \bibnamefont
  {{Mac Low}}}\ and\ \bibinfo {author} {\bibfnamefont {R.~S.}\ \bibnamefont
  {{Klessen}}},\ }\href {\doibase 10.1103/RevModPhys.76.125} {\bibfield
  {journal} {\bibinfo  {journal} {Rev. Mod. Phys.}\ }\textbf {\bibinfo {volume}
  {76}},\ \bibinfo {pages} {125} (\bibinfo {year} {2004})}\BibitemShut
  {NoStop}%
\bibitem [{\citenamefont {{Murphy}}\ \emph {et~al.}(2009)\citenamefont
  {{Murphy}}, \citenamefont {{Chary}}, \citenamefont {{Alexander}},
  \citenamefont {{Dickinson}}, \citenamefont {{Magnelli}}, \citenamefont
  {{Morrison}}, \citenamefont {{Pope}},\ and\ \citenamefont
  {{Teplitz}}}]{MurphyEtAl2009a}%
  \BibitemOpen
  \bibfield  {author} {\bibinfo {author} {\bibfnamefont {E.~J.}\ \bibnamefont
  {{Murphy}}}, \bibinfo {author} {\bibfnamefont {R.-R.}\ \bibnamefont
  {{Chary}}}, \bibinfo {author} {\bibfnamefont {D.~M.}\ \bibnamefont
  {{Alexander}}}, \bibinfo {author} {\bibfnamefont {M.}~\bibnamefont
  {{Dickinson}}}, \bibinfo {author} {\bibfnamefont {B.}~\bibnamefont
  {{Magnelli}}}, \bibinfo {author} {\bibfnamefont {G.}~\bibnamefont
  {{Morrison}}}, \bibinfo {author} {\bibfnamefont {A.}~\bibnamefont {{Pope}}},
  \ and\ \bibinfo {author} {\bibfnamefont {H.~I.}\ \bibnamefont {{Teplitz}}},\
  }\href {\doibase 10.1088/0004-637X/698/2/1380} {\bibfield  {journal}
  {\bibinfo  {journal} {\apj}\ }\textbf {\bibinfo {volume} {698}},\ \bibinfo
  {pages} {1380} (\bibinfo {year} {2009})}\BibitemShut {NoStop}%
\bibitem [{\citenamefont {{Murphy}}(2009)}]{Murphy2009}%
  \BibitemOpen
  \bibfield  {author} {\bibinfo {author} {\bibfnamefont {E.~J.}\ \bibnamefont
  {{Murphy}}},\ }\href {\doibase 10.1088/0004-637X/706/1/482} {\bibfield
  {journal} {\bibinfo  {journal} {\apj}\ }\textbf {\bibinfo {volume} {706}},\
  \bibinfo {pages} {482} (\bibinfo {year} {2009})}\BibitemShut {NoStop}%
\bibitem [{\citenamefont {{Lacki}}\ and\ \citenamefont
  {{Thompson}}(2010)}]{LackiEtAl2010b}%
  \BibitemOpen
  \bibfield  {author} {\bibinfo {author} {\bibfnamefont {B.~C.}\ \bibnamefont
  {{Lacki}}}\ and\ \bibinfo {author} {\bibfnamefont {T.~A.}\ \bibnamefont
  {{Thompson}}},\ }\href {\doibase 10.1088/0004-637X/717/1/196} {\bibfield
  {journal} {\bibinfo  {journal} {\apj}\ }\textbf {\bibinfo {volume} {717}},\
  \bibinfo {pages} {196} (\bibinfo {year} {2010})}\BibitemShut {NoStop}%
\bibitem [{\citenamefont {{Ivison}}\ \emph
  {et~al.}(2010{\natexlab{a}})\citenamefont {{Ivison}}, \citenamefont
  {{Swinbank}}, \citenamefont {{Swinyard}}, \citenamefont {{Smail}},
  \citenamefont {{Pearson}}, \citenamefont {{Rigopoulou}}, \citenamefont
  {{Polehampton}}, \citenamefont {{Baluteau}}, \citenamefont {{Barlow}},
  \citenamefont {{Blain}}, \citenamefont {{Bock}}, \citenamefont {{Clements}},
  \citenamefont {{Coppin}}, \citenamefont {{Cooray}}, \citenamefont
  {{Danielson}}, \citenamefont {{Dwek}}, \citenamefont {{Edge}}, \citenamefont
  {{Franceschini}}, \citenamefont {{Fulton}}, \citenamefont {{Glenn}},
  \citenamefont {{Griffin}}, \citenamefont {{Isaak}}, \citenamefont {{Leeks}},
  \citenamefont {{Lim}}, \citenamefont {{Naylor}}, \citenamefont {{Oliver}},
  \citenamefont {{Page}}, \citenamefont {{P{\'e}rez Fournon}}, \citenamefont
  {{Rowan-Robinson}}, \citenamefont {{Savini}}, \citenamefont {{Scott}},
  \citenamefont {{Spencer}}, \citenamefont {{Valtchanov}}, \citenamefont
  {{Vigroux}},\ and\ \citenamefont {{Wright}}}]{IvisonEtAl2010}%
  \BibitemOpen
  \bibfield  {author} {\bibinfo {author} {\bibfnamefont {R.~J.}\ \bibnamefont
  {{Ivison}}}, \bibinfo {author} {\bibfnamefont {A.~M.}\ \bibnamefont
  {{Swinbank}}}, \bibinfo {author} {\bibfnamefont {B.}~\bibnamefont
  {{Swinyard}}}, \bibinfo {author} {\bibfnamefont {I.}~\bibnamefont {{Smail}}},
  \bibinfo {author} {\bibfnamefont {C.~P.}\ \bibnamefont {{Pearson}}}, \bibinfo
  {author} {\bibfnamefont {D.}~\bibnamefont {{Rigopoulou}}}, \bibinfo {author}
  {\bibfnamefont {E.}~\bibnamefont {{Polehampton}}}, \bibinfo {author}
  {\bibfnamefont {J.-P.}\ \bibnamefont {{Baluteau}}}, \bibinfo {author}
  {\bibfnamefont {M.~J.}\ \bibnamefont {{Barlow}}}, \bibinfo {author}
  {\bibfnamefont {A.~W.}\ \bibnamefont {{Blain}}}, \bibinfo {author}
  {\bibfnamefont {J.}~\bibnamefont {{Bock}}}, \bibinfo {author} {\bibfnamefont
  {D.~L.}\ \bibnamefont {{Clements}}}, \bibinfo {author} {\bibfnamefont
  {K.}~\bibnamefont {{Coppin}}}, \bibinfo {author} {\bibfnamefont
  {A.}~\bibnamefont {{Cooray}}}, \bibinfo {author} {\bibfnamefont
  {A.}~\bibnamefont {{Danielson}}}, \bibinfo {author} {\bibfnamefont
  {E.}~\bibnamefont {{Dwek}}}, \bibinfo {author} {\bibfnamefont {A.~C.}\
  \bibnamefont {{Edge}}}, \bibinfo {author} {\bibfnamefont {A.}~\bibnamefont
  {{Franceschini}}}, \bibinfo {author} {\bibfnamefont {T.}~\bibnamefont
  {{Fulton}}}, \bibinfo {author} {\bibfnamefont {J.}~\bibnamefont {{Glenn}}},
  \bibinfo {author} {\bibfnamefont {M.}~\bibnamefont {{Griffin}}}, \bibinfo
  {author} {\bibfnamefont {K.}~\bibnamefont {{Isaak}}}, \bibinfo {author}
  {\bibfnamefont {S.}~\bibnamefont {{Leeks}}}, \bibinfo {author} {\bibfnamefont
  {T.}~\bibnamefont {{Lim}}}, \bibinfo {author} {\bibfnamefont
  {D.}~\bibnamefont {{Naylor}}}, \bibinfo {author} {\bibfnamefont {S.~J.}\
  \bibnamefont {{Oliver}}}, \bibinfo {author} {\bibfnamefont {M.~J.}\
  \bibnamefont {{Page}}}, \bibinfo {author} {\bibfnamefont {I.}~\bibnamefont
  {{P{\'e}rez Fournon}}}, \bibinfo {author} {\bibfnamefont {M.}~\bibnamefont
  {{Rowan-Robinson}}}, \bibinfo {author} {\bibfnamefont {G.}~\bibnamefont
  {{Savini}}}, \bibinfo {author} {\bibfnamefont {D.}~\bibnamefont {{Scott}}},
  \bibinfo {author} {\bibfnamefont {L.}~\bibnamefont {{Spencer}}}, \bibinfo
  {author} {\bibfnamefont {I.}~\bibnamefont {{Valtchanov}}}, \bibinfo {author}
  {\bibfnamefont {L.}~\bibnamefont {{Vigroux}}}, \ and\ \bibinfo {author}
  {\bibfnamefont {G.~S.}\ \bibnamefont {{Wright}}},\ }\href@noop {} {\bibfield
  {journal} {\bibinfo  {journal} {\aap}\ }\textbf {\bibinfo {volume} {518}}
  (\bibinfo {year} {2010}{\natexlab{a}})}\BibitemShut {NoStop}%
\bibitem [{\citenamefont {{Kennicutt}}\ and\ \citenamefont
  {{Evans}}(2012)}]{KennicuttEvans2012}%
  \BibitemOpen
  \bibfield  {author} {\bibinfo {author} {\bibfnamefont {R.~C.}\ \bibnamefont
  {{Kennicutt}}}\ and\ \bibinfo {author} {\bibfnamefont {N.~J.}\ \bibnamefont
  {{Evans}}},\ }\href {\doibase 10.1146/annurev-astro-081811-125610} {\bibfield
   {journal} {\bibinfo  {journal} {\araa}\ }\textbf {\bibinfo {volume} {50}},\
  \bibinfo {pages} {531} (\bibinfo {year} {2012})}\BibitemShut {NoStop}%
\bibitem [{\citenamefont {{Rix}}\ and\ \citenamefont
  {{Bovy}}(2013)}]{RixBovy2013}%
  \BibitemOpen
  \bibfield  {author} {\bibinfo {author} {\bibfnamefont {H.-W.}\ \bibnamefont
  {{Rix}}}\ and\ \bibinfo {author} {\bibfnamefont {J.}~\bibnamefont {{Bovy}}},\
  }\href {\doibase 10.1007/s00159-013-0061-8} {\bibfield  {journal} {\bibinfo
  {journal} {\aapr}\ }\textbf {\bibinfo {volume} {21}},\ \bibinfo {eid} {61}
  (\bibinfo {year} {2013})}\BibitemShut {NoStop}%
\bibitem [{\citenamefont {{Mo}}\ \emph {et~al.}(1998)\citenamefont {{Mo}},
  \citenamefont {{Mao}},\ and\ \citenamefont {{White}}}]{MoMaoWhite1998}%
  \BibitemOpen
  \bibfield  {author} {\bibinfo {author} {\bibfnamefont {H.~J.}\ \bibnamefont
  {{Mo}}}, \bibinfo {author} {\bibfnamefont {S.}~\bibnamefont {{Mao}}}, \ and\
  \bibinfo {author} {\bibfnamefont {S.~D.~M.}\ \bibnamefont {{White}}},\ }\href
  {\doibase 10.1046/j.1365-8711.1998.01227.x} {\bibfield  {journal} {\bibinfo
  {journal} {\mnras}\ }\textbf {\bibinfo {volume} {295}},\ \bibinfo {pages}
  {319} (\bibinfo {year} {1998})}\BibitemShut {NoStop}%
\bibitem [{\citenamefont {{Oesch}}\ \emph {et~al.}(2010)\citenamefont
  {{Oesch}}, \citenamefont {{Bouwens}}, \citenamefont {{Carollo}},
  \citenamefont {{Illingworth}}, \citenamefont {{Trenti}}, \citenamefont
  {{Stiavelli}}, \citenamefont {{Magee}}, \citenamefont {{Labb{\'e}}},\ and\
  \citenamefont {{Franx}}}]{OeschEtAl2010}%
  \BibitemOpen
  \bibfield  {author} {\bibinfo {author} {\bibfnamefont {P.~A.}\ \bibnamefont
  {{Oesch}}}, \bibinfo {author} {\bibfnamefont {R.~J.}\ \bibnamefont
  {{Bouwens}}}, \bibinfo {author} {\bibfnamefont {C.~M.}\ \bibnamefont
  {{Carollo}}}, \bibinfo {author} {\bibfnamefont {G.~D.}\ \bibnamefont
  {{Illingworth}}}, \bibinfo {author} {\bibfnamefont {M.}~\bibnamefont
  {{Trenti}}}, \bibinfo {author} {\bibfnamefont {M.}~\bibnamefont
  {{Stiavelli}}}, \bibinfo {author} {\bibfnamefont {D.}~\bibnamefont
  {{Magee}}}, \bibinfo {author} {\bibfnamefont {I.}~\bibnamefont
  {{Labb{\'e}}}}, \ and\ \bibinfo {author} {\bibfnamefont {M.}~\bibnamefont
  {{Franx}}},\ }\href {\doibase 10.1088/2041-8205/709/1/L21} {\bibfield
  {journal} {\bibinfo  {journal} {\apjl}\ }\textbf {\bibinfo {volume} {709}},\
  \bibinfo {pages} {L21} (\bibinfo {year} {2010})}\BibitemShut {NoStop}%
\bibitem [{\citenamefont {{Leroy}}\ \emph {et~al.}(2008)\citenamefont
  {{Leroy}}, \citenamefont {{Walter}}, \citenamefont {{Brinks}}, \citenamefont
  {{Bigiel}}, \citenamefont {{de Blok}}, \citenamefont {{Madore}},\ and\
  \citenamefont {{Thornley}}}]{LeroyEtAl2008}%
  \BibitemOpen
  \bibfield  {author} {\bibinfo {author} {\bibfnamefont {A.~K.}\ \bibnamefont
  {{Leroy}}}, \bibinfo {author} {\bibfnamefont {F.}~\bibnamefont {{Walter}}},
  \bibinfo {author} {\bibfnamefont {E.}~\bibnamefont {{Brinks}}}, \bibinfo
  {author} {\bibfnamefont {F.}~\bibnamefont {{Bigiel}}}, \bibinfo {author}
  {\bibfnamefont {W.~J.~G.}\ \bibnamefont {{de Blok}}}, \bibinfo {author}
  {\bibfnamefont {B.}~\bibnamefont {{Madore}}}, \ and\ \bibinfo {author}
  {\bibfnamefont {M.~D.}\ \bibnamefont {{Thornley}}},\ }\href {\doibase
  10.1088/0004-6256/136/6/2782} {\bibfield  {journal} {\bibinfo  {journal}
  {\aj}\ }\textbf {\bibinfo {volume} {136}},\ \bibinfo {pages} {2782} (\bibinfo
  {year} {2008})}\BibitemShut {NoStop}%
\bibitem [{\citenamefont {{Kroupa}}(2002)}]{Kroupa2002}%
  \BibitemOpen
  \bibfield  {author} {\bibinfo {author} {\bibfnamefont {P.}~\bibnamefont
  {{Kroupa}}},\ }\href {\doibase 10.1126/science.1067524} {\bibfield  {journal}
  {\bibinfo  {journal} {Science}\ }\textbf {\bibinfo {volume} {295}},\ \bibinfo
  {pages} {82} (\bibinfo {year} {2002})}\BibitemShut {NoStop}%
\bibitem [{\citenamefont {{Cirelli}}\ and\ \citenamefont
  {{Panci}}(2009)}]{CirelliPanci2009}%
  \BibitemOpen
  \bibfield  {author} {\bibinfo {author} {\bibfnamefont {M.}~\bibnamefont
  {{Cirelli}}}\ and\ \bibinfo {author} {\bibfnamefont {P.}~\bibnamefont
  {{Panci}}},\ }\href {\doibase 10.1016/j.nuclphysb.2009.06.034} {\bibfield
  {journal} {\bibinfo  {journal} {Nuclear Physics B}\ }\textbf {\bibinfo
  {volume} {821}},\ \bibinfo {pages} {399} (\bibinfo {year}
  {2009})}\BibitemShut {NoStop}%
\bibitem [{\citenamefont {{Chakraborty}}\ and\ \citenamefont
  {{Fields}}(2013)}]{ChakrabortyFields2013}%
  \BibitemOpen
  \bibfield  {author} {\bibinfo {author} {\bibfnamefont {N.}~\bibnamefont
  {{Chakraborty}}}\ and\ \bibinfo {author} {\bibfnamefont {B.~D.}\ \bibnamefont
  {{Fields}}},\ }\href {\doibase 10.1088/0004-637X/773/2/104} {\bibfield
  {journal} {\bibinfo  {journal} {\apj}\ }\textbf {\bibinfo {volume} {773}},\
  \bibinfo {eid} {104} (\bibinfo {year} {2013})}\BibitemShut {NoStop}%
\bibitem [{\citenamefont {{Galliano}}\ \emph {et~al.}(2008)\citenamefont
  {{Galliano}}, \citenamefont {{Dwek}},\ and\ \citenamefont
  {{Chanial}}}]{GallianoEtAl2008}%
  \BibitemOpen
  \bibfield  {author} {\bibinfo {author} {\bibfnamefont {F.}~\bibnamefont
  {{Galliano}}}, \bibinfo {author} {\bibfnamefont {E.}~\bibnamefont {{Dwek}}},
  \ and\ \bibinfo {author} {\bibfnamefont {P.}~\bibnamefont {{Chanial}}},\
  }\href {\doibase 10.1086/523621} {\bibfield  {journal} {\bibinfo  {journal}
  {\apj}\ }\textbf {\bibinfo {volume} {672}},\ \bibinfo {eid} {214} (\bibinfo
  {year} {2008})}\BibitemShut {NoStop}%
\bibitem [{\citenamefont {{Girichidis}}\ \emph {et~al.}(2015)\citenamefont
  {{Girichidis}}, \citenamefont {{Naab}}, \citenamefont {{Walch}},
  \citenamefont {{Hanasz}}, \citenamefont {{Mac Low}}, \citenamefont
  {{Ostriker}}, \citenamefont {{Gatto}}, \citenamefont {{Peters}},
  \citenamefont {{W{\"u}nsch}}, \citenamefont {{Glover}}, \citenamefont
  {{Klessen}}, \citenamefont {{Clark}},\ and\ \citenamefont
  {{Baczynski}}}]{GirichidisEtAl2015}%
  \BibitemOpen
  \bibfield  {author} {\bibinfo {author} {\bibfnamefont {P.}~\bibnamefont
  {{Girichidis}}}, \bibinfo {author} {\bibfnamefont {T.}~\bibnamefont
  {{Naab}}}, \bibinfo {author} {\bibfnamefont {S.}~\bibnamefont {{Walch}}},
  \bibinfo {author} {\bibfnamefont {M.}~\bibnamefont {{Hanasz}}}, \bibinfo
  {author} {\bibfnamefont {M.-M.}\ \bibnamefont {{Mac Low}}}, \bibinfo {author}
  {\bibfnamefont {J.~P.}\ \bibnamefont {{Ostriker}}}, \bibinfo {author}
  {\bibfnamefont {A.}~\bibnamefont {{Gatto}}}, \bibinfo {author} {\bibfnamefont
  {T.}~\bibnamefont {{Peters}}}, \bibinfo {author} {\bibfnamefont
  {R.}~\bibnamefont {{W{\"u}nsch}}}, \bibinfo {author} {\bibfnamefont
  {S.~C.~O.}\ \bibnamefont {{Glover}}}, \bibinfo {author} {\bibfnamefont
  {R.~S.}\ \bibnamefont {{Klessen}}}, \bibinfo {author} {\bibfnamefont {P.~C.}\
  \bibnamefont {{Clark}}}, \ and\ \bibinfo {author} {\bibfnamefont
  {C.}~\bibnamefont {{Baczynski}}},\ }\href@noop {} {\bibfield  {journal}
  {\bibinfo  {journal} {ArXiv e-prints}\ } (\bibinfo {year} {2015})},\ \Eprint
  {http://arxiv.org/abs/1509.07247} {arXiv:1509.07247} \BibitemShut {NoStop}%
\bibitem [{\citenamefont {{Shu}}\ \emph {et~al.}(2005)\citenamefont {{Shu}},
  \citenamefont {{Mo}},\ and\ \citenamefont
  {{Shu-DeMao}}}]{ShuMoShu-DeMao2005}%
  \BibitemOpen
  \bibfield  {author} {\bibinfo {author} {\bibfnamefont {C.-G.}\ \bibnamefont
  {{Shu}}}, \bibinfo {author} {\bibfnamefont {H.-J.}\ \bibnamefont {{Mo}}}, \
  and\ \bibinfo {author} {\bibnamefont {{Shu-DeMao}}},\ }\href {\doibase
  10.1088/1009-9271/5/4/001} {\bibfield  {journal} {\bibinfo  {journal}
  {\cjaa}\ }\textbf {\bibinfo {volume} {5}},\ \bibinfo {pages} {327} (\bibinfo
  {year} {2005})}\BibitemShut {NoStop}%
\bibitem [{\citenamefont {{Girichidis}}\ \emph {et~al.}(2016)\citenamefont
  {{Girichidis}}, \citenamefont {{Walch}}, \citenamefont {{Naab}},
  \citenamefont {{Gatto}}, \citenamefont {{W{\"u}nsch}}, \citenamefont
  {{Glover}}, \citenamefont {{Klessen}}, \citenamefont {{Clark}}, \citenamefont
  {{Peters}}, \citenamefont {{Derigs}},\ and\ \citenamefont
  {{Baczynski}}}]{GirichidisEtAl2016}%
  \BibitemOpen
  \bibfield  {author} {\bibinfo {author} {\bibfnamefont {P.}~\bibnamefont
  {{Girichidis}}}, \bibinfo {author} {\bibfnamefont {S.}~\bibnamefont
  {{Walch}}}, \bibinfo {author} {\bibfnamefont {T.}~\bibnamefont {{Naab}}},
  \bibinfo {author} {\bibfnamefont {A.}~\bibnamefont {{Gatto}}}, \bibinfo
  {author} {\bibfnamefont {R.}~\bibnamefont {{W{\"u}nsch}}}, \bibinfo {author}
  {\bibfnamefont {S.~C.~O.}\ \bibnamefont {{Glover}}}, \bibinfo {author}
  {\bibfnamefont {R.~S.}\ \bibnamefont {{Klessen}}}, \bibinfo {author}
  {\bibfnamefont {P.~C.}\ \bibnamefont {{Clark}}}, \bibinfo {author}
  {\bibfnamefont {T.}~\bibnamefont {{Peters}}}, \bibinfo {author}
  {\bibfnamefont {D.}~\bibnamefont {{Derigs}}}, \ and\ \bibinfo {author}
  {\bibfnamefont {C.}~\bibnamefont {{Baczynski}}},\ }\href {\doibase
  10.1093/mnras/stv2742} {\bibfield  {journal} {\bibinfo  {journal} {\mnras}\
  }\textbf {\bibinfo {volume} {456}},\ \bibinfo {pages} {3432} (\bibinfo {year}
  {2016})}\BibitemShut {NoStop}%
\bibitem [{\citenamefont {{Walter}}\ \emph {et~al.}(2002)\citenamefont
  {{Walter}}, \citenamefont {{Weiss}},\ and\ \citenamefont
  {{Scoville}}}]{WalterEtAl2002}%
  \BibitemOpen
  \bibfield  {author} {\bibinfo {author} {\bibfnamefont {F.}~\bibnamefont
  {{Walter}}}, \bibinfo {author} {\bibfnamefont {A.}~\bibnamefont {{Weiss}}}, \
  and\ \bibinfo {author} {\bibfnamefont {N.}~\bibnamefont {{Scoville}}},\
  }\href {\doibase 10.1086/345287} {\bibfield  {journal} {\bibinfo  {journal}
  {\apjl}\ }\textbf {\bibinfo {volume} {580}},\ \bibinfo {pages} {L21}
  (\bibinfo {year} {2002})}\BibitemShut {NoStop}%
\bibitem [{\citenamefont {{Kazantsev}}(1968)}]{Kazantsev1968}%
  \BibitemOpen
  \bibfield  {author} {\bibinfo {author} {\bibfnamefont {A.~P.}\ \bibnamefont
  {{Kazantsev}}},\ }\href@noop {} {\bibfield  {journal} {\bibinfo  {journal}
  {Soviet Journal of Experimental and Theoretical Physics}\ }\textbf {\bibinfo
  {volume} {26}},\ \bibinfo {pages} {1031} (\bibinfo {year}
  {1968})}\BibitemShut {NoStop}%
\bibitem [{\citenamefont {{Kulsrud}}\ and\ \citenamefont
  {{Anderson}}(1992)}]{KulsrudAnderson1992}%
  \BibitemOpen
  \bibfield  {author} {\bibinfo {author} {\bibfnamefont {R.~M.}\ \bibnamefont
  {{Kulsrud}}}\ and\ \bibinfo {author} {\bibfnamefont {S.~W.}\ \bibnamefont
  {{Anderson}}},\ }\href {\doibase 10.1086/171743} {\bibfield  {journal}
  {\bibinfo  {journal} {\apj}\ }\textbf {\bibinfo {volume} {396}},\ \bibinfo
  {pages} {606} (\bibinfo {year} {1992})}\BibitemShut {NoStop}%
\bibitem [{\citenamefont {{Schober}}\ \emph
  {et~al.}(2012{\natexlab{c}})\citenamefont {{Schober}}, \citenamefont
  {{Schleicher}}, \citenamefont {{Federrath}}, \citenamefont {{Glover}},
  \citenamefont {{Klessen}},\ and\ \citenamefont
  {{Banerjee}}}]{SchoberEtAl2012.2}%
  \BibitemOpen
  \bibfield  {author} {\bibinfo {author} {\bibfnamefont {J.}~\bibnamefont
  {{Schober}}}, \bibinfo {author} {\bibfnamefont {D.}~\bibnamefont
  {{Schleicher}}}, \bibinfo {author} {\bibfnamefont {C.}~\bibnamefont
  {{Federrath}}}, \bibinfo {author} {\bibfnamefont {S.}~\bibnamefont
  {{Glover}}}, \bibinfo {author} {\bibfnamefont {R.~S.}\ \bibnamefont
  {{Klessen}}}, \ and\ \bibinfo {author} {\bibfnamefont {R.}~\bibnamefont
  {{Banerjee}}},\ }\href {\doibase 10.1088/0004-637X/754/2/99} {\bibfield
  {journal} {\bibinfo  {journal} {\apj}\ }\textbf {\bibinfo {volume} {754}},\
  \bibinfo {eid} {99} (\bibinfo {year} {2012}{\natexlab{c}})}\BibitemShut
  {NoStop}%
\bibitem [{\citenamefont {{Pakmor}}\ and\ \citenamefont
  {{Springel}}(2013)}]{PakmorSpringel2013}%
  \BibitemOpen
  \bibfield  {author} {\bibinfo {author} {\bibfnamefont {R.}~\bibnamefont
  {{Pakmor}}}\ and\ \bibinfo {author} {\bibfnamefont {V.}~\bibnamefont
  {{Springel}}},\ }\href {\doibase 10.1093/mnras/stt428} {\bibfield  {journal}
  {\bibinfo  {journal} {\mnras}\ }\textbf {\bibinfo {volume} {432}},\ \bibinfo
  {pages} {176} (\bibinfo {year} {2013})}\BibitemShut {NoStop}%
\bibitem [{\citenamefont {{Schober}}\ \emph {et~al.}(2015)\citenamefont
  {{Schober}}, \citenamefont {{Schleicher}}, \citenamefont {{Federrath}},
  \citenamefont {{Bovino}},\ and\ \citenamefont {{Klessen}}}]{SchoberEtAl2015}%
  \BibitemOpen
  \bibfield  {author} {\bibinfo {author} {\bibfnamefont {J.}~\bibnamefont
  {{Schober}}}, \bibinfo {author} {\bibfnamefont {D.~R.~G.}\ \bibnamefont
  {{Schleicher}}}, \bibinfo {author} {\bibfnamefont {C.}~\bibnamefont
  {{Federrath}}}, \bibinfo {author} {\bibfnamefont {S.}~\bibnamefont
  {{Bovino}}}, \ and\ \bibinfo {author} {\bibfnamefont {R.~S.}\ \bibnamefont
  {{Klessen}}},\ }\href {\doibase 10.1103/PhysRevE.92.023010} {\bibfield
  {journal} {\bibinfo  {journal} {\pre}\ }\textbf {\bibinfo {volume} {92}},\
  \bibinfo {eid} {023010} (\bibinfo {year} {2015})}\BibitemShut {NoStop}%
\bibitem [{\citenamefont {{Gazol-Pati{\~n}o}}\ and\ \citenamefont
  {{Passot}}(1999)}]{Gazol-PatinoPassot1999}%
  \BibitemOpen
  \bibfield  {author} {\bibinfo {author} {\bibfnamefont {A.}~\bibnamefont
  {{Gazol-Pati{\~n}o}}}\ and\ \bibinfo {author} {\bibfnamefont
  {T.}~\bibnamefont {{Passot}}},\ }\href {\doibase 10.1086/307306} {\bibfield
  {journal} {\bibinfo  {journal} {\apj}\ }\textbf {\bibinfo {volume} {518}},\
  \bibinfo {pages} {748} (\bibinfo {year} {1999})}\BibitemShut {NoStop}%
\bibitem [{\citenamefont {{Korpi}}\ \emph {et~al.}(1998)\citenamefont
  {{Korpi}}, \citenamefont {{Brandenburg}},\ and\ \citenamefont
  {{Touminen}}}]{KorpiBrandenburgTouminen1998}%
  \BibitemOpen
  \bibfield  {author} {\bibinfo {author} {\bibfnamefont {M.}~\bibnamefont
  {{Korpi}}}, \bibinfo {author} {\bibfnamefont {A.}~\bibnamefont
  {{Brandenburg}}}, \ and\ \bibinfo {author} {\bibfnamefont {I.}~\bibnamefont
  {{Touminen}}},\ }\href@noop {} {\bibfield  {journal} {\bibinfo  {journal}
  {Studia geophys.~et geod.}\ }\textbf {\bibinfo {volume} {42}},\ \bibinfo
  {pages} {410} (\bibinfo {year} {1998})}\BibitemShut {NoStop}%
\bibitem [{\citenamefont {{Klessen}}\ and\ \citenamefont
  {{Hennebelle}}(2010)}]{KlessenHennebelle2010}%
  \BibitemOpen
  \bibfield  {author} {\bibinfo {author} {\bibfnamefont {R.~S.}\ \bibnamefont
  {{Klessen}}}\ and\ \bibinfo {author} {\bibfnamefont {P.}~\bibnamefont
  {{Hennebelle}}},\ }\href {\doibase 10.1051/0004-6361/200913780} {\bibfield
  {journal} {\bibinfo  {journal} {\aap}\ }\textbf {\bibinfo {volume} {520}},\
  \bibinfo {pages} {A17} (\bibinfo {year} {2010})}\BibitemShut {NoStop}%
\bibitem [{\citenamefont {{Klessen}}\ and\ \citenamefont
  {{Glover}}(2016)}]{KlessenGlover2016}%
  \BibitemOpen
  \bibfield  {author} {\bibinfo {author} {\bibfnamefont {R.~S.}\ \bibnamefont
  {{Klessen}}}\ and\ \bibinfo {author} {\bibfnamefont {S.~C.~O.}\ \bibnamefont
  {{Glover}}},\ }\href {\doibase 10.1007/978-3-662-47890-5_2} {\bibfield
  {journal} {\bibinfo  {journal} {Saas-Fee Advanced Course, Springer-Verlag
  Berlin Heidelberg}\ }\textbf {\bibinfo {volume} {43}},\ \bibinfo {pages} {85}
  (\bibinfo {year} {2016})}\BibitemShut {NoStop}%
\bibitem [{\citenamefont {{Niklas}}\ and\ \citenamefont
  {{Beck}}(1997)}]{NiklasBeck1997}%
  \BibitemOpen
  \bibfield  {author} {\bibinfo {author} {\bibfnamefont {S.}~\bibnamefont
  {{Niklas}}}\ and\ \bibinfo {author} {\bibfnamefont {R.}~\bibnamefont
  {{Beck}}},\ }\href@noop {} {\bibfield  {journal} {\bibinfo  {journal} {\aap}\
  }\textbf {\bibinfo {volume} {320}},\ \bibinfo {pages} {54} (\bibinfo {year}
  {1997})}\BibitemShut {NoStop}%
\bibitem [{\citenamefont {{Chy{\.z}y}}\ \emph {et~al.}(2011)\citenamefont
  {{Chy{\.z}y}}, \citenamefont {{We{\.z}gowiec}}, \citenamefont {{Beck}},\ and\
  \citenamefont {{Bomans}}}]{ChyzyEtAl2011}%
  \BibitemOpen
  \bibfield  {author} {\bibinfo {author} {\bibfnamefont {K.~T.}\ \bibnamefont
  {{Chy{\.z}y}}}, \bibinfo {author} {\bibfnamefont {M.}~\bibnamefont
  {{We{\.z}gowiec}}}, \bibinfo {author} {\bibfnamefont {R.}~\bibnamefont
  {{Beck}}}, \ and\ \bibinfo {author} {\bibfnamefont {D.~J.}\ \bibnamefont
  {{Bomans}}},\ }\href {\doibase 10.1051/0004-6361/201015393} {\bibfield
  {journal} {\bibinfo  {journal} {\aap}\ }\textbf {\bibinfo {volume} {529}},\
  \bibinfo {eid} {A94} (\bibinfo {year} {2011})}\BibitemShut {NoStop}%
\bibitem [{\citenamefont {{Adebahr}}\ \emph {et~al.}(2013)\citenamefont
  {{Adebahr}}, \citenamefont {{Krause}}, \citenamefont {{Klein}}, \citenamefont
  {{We{\.z}gowiec}}, \citenamefont {{Bomans}},\ and\ \citenamefont
  {{Dettmar}}}]{AdebahrEtAl2013}%
  \BibitemOpen
  \bibfield  {author} {\bibinfo {author} {\bibfnamefont {B.}~\bibnamefont
  {{Adebahr}}}, \bibinfo {author} {\bibfnamefont {M.}~\bibnamefont {{Krause}}},
  \bibinfo {author} {\bibfnamefont {U.}~\bibnamefont {{Klein}}}, \bibinfo
  {author} {\bibfnamefont {M.}~\bibnamefont {{We{\.z}gowiec}}}, \bibinfo
  {author} {\bibfnamefont {D.~J.}\ \bibnamefont {{Bomans}}}, \ and\ \bibinfo
  {author} {\bibfnamefont {R.-J.}\ \bibnamefont {{Dettmar}}},\ }\href {\doibase
  10.1051/0004-6361/201220226} {\bibfield  {journal} {\bibinfo  {journal}
  {\aap}\ }\textbf {\bibinfo {volume} {555}},\ \bibinfo {eid} {A23} (\bibinfo
  {year} {2013})}\BibitemShut {NoStop}%
\bibitem [{\citenamefont {{Bell}}(1978{\natexlab{a}})}]{Bell1978a}%
  \BibitemOpen
  \bibfield  {author} {\bibinfo {author} {\bibfnamefont {A.~R.}\ \bibnamefont
  {{Bell}}},\ }\href@noop {} {\bibfield  {journal} {\bibinfo  {journal}
  {\mnras}\ }\textbf {\bibinfo {volume} {182}},\ \bibinfo {pages} {147}
  (\bibinfo {year} {1978}{\natexlab{a}})}\BibitemShut {NoStop}%
\bibitem [{\citenamefont {{Bell}}(1978{\natexlab{b}})}]{Bell1978b}%
  \BibitemOpen
  \bibfield  {author} {\bibinfo {author} {\bibfnamefont {A.~R.}\ \bibnamefont
  {{Bell}}},\ }\href@noop {} {\bibfield  {journal} {\bibinfo  {journal}
  {\mnras}\ }\textbf {\bibinfo {volume} {182}},\ \bibinfo {pages} {443}
  (\bibinfo {year} {1978}{\natexlab{b}})}\BibitemShut {NoStop}%
\bibitem [{\citenamefont {{Drury}}(1983)}]{Drury1983}%
  \BibitemOpen
  \bibfield  {author} {\bibinfo {author} {\bibfnamefont {L.}~\bibnamefont
  {{Drury}}},\ }\href {\doibase 10.1007/BF00171901} {\bibfield  {journal}
  {\bibinfo  {journal} {\ssr}\ }\textbf {\bibinfo {volume} {36}},\ \bibinfo
  {pages} {57} (\bibinfo {year} {1983})}\BibitemShut {NoStop}%
\bibitem [{\citenamefont {{Schlickeiser}}(2002)}]{Schlickeiser2002}%
  \BibitemOpen
  \bibfield  {author} {\bibinfo {author} {\bibfnamefont {R.}~\bibnamefont
  {{Schlickeiser}}},\ }\href@noop {} {\emph {\bibinfo {title} {Cosmic ray
  astrophysics / Reinhard Schlickeiser, Astronomy and Astrophysics Library;
  Physics and Astronomy Online Library.~Berlin: Springer.~ISBN 3-540-66465-3,
  2002, XV + 519 pp.}}}\ (\bibinfo {year} {2002})\BibitemShut {NoStop}%
\bibitem [{\citenamefont {{Bogdan}}\ and\ \citenamefont
  {{V{\"o}lk}}(1983)}]{BogdanVolk1983}%
  \BibitemOpen
  \bibfield  {author} {\bibinfo {author} {\bibfnamefont {T.~J.}\ \bibnamefont
  {{Bogdan}}}\ and\ \bibinfo {author} {\bibfnamefont {H.~J.}\ \bibnamefont
  {{V{\"o}lk}}},\ }\href@noop {} {\bibfield  {journal} {\bibinfo  {journal}
  {\aap}\ }\textbf {\bibinfo {volume} {122}},\ \bibinfo {pages} {129} (\bibinfo
  {year} {1983})}\BibitemShut {NoStop}%
\bibitem [{\citenamefont {{Dorfi}}(2000)}]{Dorfi2000}%
  \BibitemOpen
  \bibfield  {author} {\bibinfo {author} {\bibfnamefont {E.~A.}\ \bibnamefont
  {{Dorfi}}},\ }\href {\doibase 10.1023/A:1002648630489} {\bibfield  {journal}
  {\bibinfo  {journal} {\apss}\ }\textbf {\bibinfo {volume} {272}},\ \bibinfo
  {pages} {227} (\bibinfo {year} {2000})}\BibitemShut {NoStop}%
\bibitem [{\citenamefont {{Longair}}(2011)}]{Longair2011}%
  \BibitemOpen
  \bibfield  {author} {\bibinfo {author} {\bibfnamefont {M.~S.}\ \bibnamefont
  {{Longair}}},\ }\href@noop {} {\emph {\bibinfo {title} {High Energy
  Astrophysics, by Malcolm S.~Longair, Cambridge, UK: Cambridge University
  Press, 2011}}}\ (\bibinfo {year} {2011})\BibitemShut {NoStop}%
\bibitem [{\citenamefont {{Strong}}\ and\ \citenamefont
  {{Moskalenko}}(1998)}]{StrongMoskalenko1998}%
  \BibitemOpen
  \bibfield  {author} {\bibinfo {author} {\bibfnamefont {A.~W.}\ \bibnamefont
  {{Strong}}}\ and\ \bibinfo {author} {\bibfnamefont {I.~V.}\ \bibnamefont
  {{Moskalenko}}},\ }\href {\doibase 10.1086/306470} {\bibfield  {journal}
  {\bibinfo  {journal} {\apj}\ }\textbf {\bibinfo {volume} {509}},\ \bibinfo
  {pages} {212} (\bibinfo {year} {1998})}\BibitemShut {NoStop}%
\bibitem [{\citenamefont {{Lacki}}\ \emph {et~al.}(2011)\citenamefont
  {{Lacki}}, \citenamefont {{Thompson}}, \citenamefont {{Quataert}},
  \citenamefont {{Loeb}},\ and\ \citenamefont {{Waxman}}}]{LackiEtAl2011}%
  \BibitemOpen
  \bibfield  {author} {\bibinfo {author} {\bibfnamefont {B.~C.}\ \bibnamefont
  {{Lacki}}}, \bibinfo {author} {\bibfnamefont {T.~A.}\ \bibnamefont
  {{Thompson}}}, \bibinfo {author} {\bibfnamefont {E.}~\bibnamefont
  {{Quataert}}}, \bibinfo {author} {\bibfnamefont {A.}~\bibnamefont {{Loeb}}},
  \ and\ \bibinfo {author} {\bibfnamefont {E.}~\bibnamefont {{Waxman}}},\
  }\href {\doibase 10.1088/0004-637X/734/2/107} {\bibfield  {journal} {\bibinfo
   {journal} {\apj}\ }\textbf {\bibinfo {volume} {734}},\ \bibinfo {eid} {107}
  (\bibinfo {year} {2011})}\BibitemShut {NoStop}%
\bibitem [{\citenamefont {{Mannheim}}\ and\ \citenamefont
  {{Schlickeiser}}(1994)}]{MannheimSchlickeiser1994}%
  \BibitemOpen
  \bibfield  {author} {\bibinfo {author} {\bibfnamefont {K.}~\bibnamefont
  {{Mannheim}}}\ and\ \bibinfo {author} {\bibfnamefont {R.}~\bibnamefont
  {{Schlickeiser}}},\ }\href@noop {} {\bibfield  {journal} {\bibinfo  {journal}
  {\aap}\ }\textbf {\bibinfo {volume} {286}} (\bibinfo {year}
  {1994})}\BibitemShut {NoStop}%
\bibitem [{\citenamefont {{Lacki}}\ and\ \citenamefont
  {{Beck}}(2013)}]{LackiBeck2013}%
  \BibitemOpen
  \bibfield  {author} {\bibinfo {author} {\bibfnamefont {B.~C.}\ \bibnamefont
  {{Lacki}}}\ and\ \bibinfo {author} {\bibfnamefont {R.}~\bibnamefont
  {{Beck}}},\ }\href {\doibase 10.1093/mnras/stt122} {\bibfield  {journal}
  {\bibinfo  {journal} {\mnras}\ }\textbf {\bibinfo {volume} {430}},\ \bibinfo
  {pages} {3171} (\bibinfo {year} {2013})}\BibitemShut {NoStop}%
\bibitem [{\citenamefont {{Blumenthal}}\ and\ \citenamefont
  {{Gould}}(1970)}]{BlumenthalGould1970}%
  \BibitemOpen
  \bibfield  {author} {\bibinfo {author} {\bibfnamefont {G.~R.}\ \bibnamefont
  {{Blumenthal}}}\ and\ \bibinfo {author} {\bibfnamefont {R.~J.}\ \bibnamefont
  {{Gould}}},\ }\href {\doibase 10.1103/RevModPhys.42.237} {\bibfield
  {journal} {\bibinfo  {journal} {Reviews of Modern Physics}\ }\textbf
  {\bibinfo {volume} {42}},\ \bibinfo {pages} {237} (\bibinfo {year}
  {1970})}\BibitemShut {NoStop}%
\bibitem [{\citenamefont {{Tielens}}(2005)}]{Tielens2005}%
  \BibitemOpen
  \bibfield  {author} {\bibinfo {author} {\bibfnamefont {A.~G.~G.~M.}\
  \bibnamefont {{Tielens}}},\ }\href@noop {} {\emph {\bibinfo {title} {The
  Physics and Chemistry of the Interstellar Medium, by A.~G.~G.~M.~Tielens,
  pp.~.~ISBN 0521826349.~Cambridge, UK: Cambridge University Press, 2005.}}}\
  (\bibinfo {year} {2005})\BibitemShut {NoStop}%
\bibitem [{\citenamefont {{Ehle}}\ and\ \citenamefont
  {{Beck}}(1993)}]{EhleBeck1993}%
  \BibitemOpen
  \bibfield  {author} {\bibinfo {author} {\bibfnamefont {M.}~\bibnamefont
  {{Ehle}}}\ and\ \bibinfo {author} {\bibfnamefont {R.}~\bibnamefont
  {{Beck}}},\ }\href@noop {} {\bibfield  {journal} {\bibinfo  {journal} {\aap}\
  }\textbf {\bibinfo {volume} {273}},\ \bibinfo {pages} {45} (\bibinfo {year}
  {1993})}\BibitemShut {NoStop}%
\bibitem [{\citenamefont {{Berkhuijsen}}\ \emph {et~al.}(2006)\citenamefont
  {{Berkhuijsen}}, \citenamefont {{Mitra}},\ and\ \citenamefont
  {{Mueller}}}]{BerkhuijsenEtAl2006}%
  \BibitemOpen
  \bibfield  {author} {\bibinfo {author} {\bibfnamefont {E.~M.}\ \bibnamefont
  {{Berkhuijsen}}}, \bibinfo {author} {\bibfnamefont {D.}~\bibnamefont
  {{Mitra}}}, \ and\ \bibinfo {author} {\bibfnamefont {P.}~\bibnamefont
  {{Mueller}}},\ }\href {\doibase 10.1002/asna.200510488} {\bibfield  {journal}
  {\bibinfo  {journal} {Astronomische Nachrichten}\ }\textbf {\bibinfo {volume}
  {327}},\ \bibinfo {pages} {82} (\bibinfo {year} {2006})}\BibitemShut
  {NoStop}%
\bibitem [{\citenamefont {{Beck}}(2007)}]{Beck2007}%
  \BibitemOpen
  \bibfield  {author} {\bibinfo {author} {\bibfnamefont {R.}~\bibnamefont
  {{Beck}}},\ }\href {\doibase 10.1051/0004-6361:20066988} {\bibfield
  {journal} {\bibinfo  {journal} {\aap}\ }\textbf {\bibinfo {volume} {470}},\
  \bibinfo {pages} {539} (\bibinfo {year} {2007})}\BibitemShut {NoStop}%
\bibitem [{\citenamefont {{Leitherer}}\ and\ \citenamefont
  {{Heckman}}(1995)}]{LeithererHeckman1995}%
  \BibitemOpen
  \bibfield  {author} {\bibinfo {author} {\bibfnamefont {C.}~\bibnamefont
  {{Leitherer}}}\ and\ \bibinfo {author} {\bibfnamefont {T.~M.}\ \bibnamefont
  {{Heckman}}},\ }\href {\doibase 10.1086/192112} {\bibfield  {journal}
  {\bibinfo  {journal} {\apjs}\ }\textbf {\bibinfo {volume} {96}},\ \bibinfo
  {pages} {9} (\bibinfo {year} {1995})}\BibitemShut {NoStop}%
\bibitem [{\citenamefont {{Salpeter}}(1955)}]{Salpeter1955}%
  \BibitemOpen
  \bibfield  {author} {\bibinfo {author} {\bibfnamefont {E.~E.}\ \bibnamefont
  {{Salpeter}}},\ }\href@noop {} {\bibfield  {journal} {\bibinfo  {journal}
  {\apj}\ }\textbf {\bibinfo {volume} {121}},\ \bibinfo {pages} {161} (\bibinfo
  {year} {1955})}\BibitemShut {NoStop}%
\bibitem [{\citenamefont {{Chapman}}\ \emph {et~al.}(2000)\citenamefont
  {{Chapman}}, \citenamefont {{Scott}}, \citenamefont {{Steidel}},
  \citenamefont {{Borys}}, \citenamefont {{Halpern}}, \citenamefont {{Morris}},
  \citenamefont {{Adelberger}}, \citenamefont {{Dickinson}}, \citenamefont
  {{Giavalisco}},\ and\ \citenamefont {{Pettini}}}]{ChapmanEtAl2000}%
  \BibitemOpen
  \bibfield  {author} {\bibinfo {author} {\bibfnamefont {S.~C.}\ \bibnamefont
  {{Chapman}}}, \bibinfo {author} {\bibfnamefont {D.}~\bibnamefont {{Scott}}},
  \bibinfo {author} {\bibfnamefont {C.~C.}\ \bibnamefont {{Steidel}}}, \bibinfo
  {author} {\bibfnamefont {C.}~\bibnamefont {{Borys}}}, \bibinfo {author}
  {\bibfnamefont {M.}~\bibnamefont {{Halpern}}}, \bibinfo {author}
  {\bibfnamefont {S.~L.}\ \bibnamefont {{Morris}}}, \bibinfo {author}
  {\bibfnamefont {K.~L.}\ \bibnamefont {{Adelberger}}}, \bibinfo {author}
  {\bibfnamefont {M.}~\bibnamefont {{Dickinson}}}, \bibinfo {author}
  {\bibfnamefont {M.}~\bibnamefont {{Giavalisco}}}, \ and\ \bibinfo {author}
  {\bibfnamefont {M.}~\bibnamefont {{Pettini}}},\ }\href {\doibase
  10.1046/j.1365-8711.2000.03866.x} {\bibfield  {journal} {\bibinfo  {journal}
  {\mnras}\ }\textbf {\bibinfo {volume} {319}},\ \bibinfo {pages} {318}
  (\bibinfo {year} {2000})}\BibitemShut {NoStop}%
\bibitem [{\citenamefont {{Helou}}\ \emph {et~al.}(1985)\citenamefont
  {{Helou}}, \citenamefont {{Soifer}},\ and\ \citenamefont
  {{Rowan-Robinson}}}]{HelouEtAl1985}%
  \BibitemOpen
  \bibfield  {author} {\bibinfo {author} {\bibfnamefont {G.}~\bibnamefont
  {{Helou}}}, \bibinfo {author} {\bibfnamefont {B.~T.}\ \bibnamefont
  {{Soifer}}}, \ and\ \bibinfo {author} {\bibfnamefont {M.}~\bibnamefont
  {{Rowan-Robinson}}},\ }\href {\doibase 10.1086/184556} {\bibfield  {journal}
  {\bibinfo  {journal} {\apjl}\ }\textbf {\bibinfo {volume} {298}},\ \bibinfo
  {pages} {L7} (\bibinfo {year} {1985})}\BibitemShut {NoStop}%
\bibitem [{\citenamefont {{Barger}}\ \emph {et~al.}(2012)\citenamefont
  {{Barger}}, \citenamefont {{Wang}}, \citenamefont {{Cowie}}, \citenamefont
  {{Owen}}, \citenamefont {{Chen}},\ and\ \citenamefont
  {{Williams}}}]{BargerEtAl2012}%
  \BibitemOpen
  \bibfield  {author} {\bibinfo {author} {\bibfnamefont {A.~J.}\ \bibnamefont
  {{Barger}}}, \bibinfo {author} {\bibfnamefont {W.-H.}\ \bibnamefont
  {{Wang}}}, \bibinfo {author} {\bibfnamefont {L.~L.}\ \bibnamefont {{Cowie}}},
  \bibinfo {author} {\bibfnamefont {F.~N.}\ \bibnamefont {{Owen}}}, \bibinfo
  {author} {\bibfnamefont {C.-C.}\ \bibnamefont {{Chen}}}, \ and\ \bibinfo
  {author} {\bibfnamefont {J.~P.}\ \bibnamefont {{Williams}}},\ }\href
  {\doibase 10.1088/0004-637X/761/2/89} {\bibfield  {journal} {\bibinfo
  {journal} {\apj}\ }\textbf {\bibinfo {volume} {761}},\ \bibinfo {eid} {89}
  (\bibinfo {year} {2012})}\BibitemShut {NoStop}%
\bibitem [{\citenamefont {{Ivison}}\ \emph
  {et~al.}(2010{\natexlab{b}})\citenamefont {{Ivison}}, \citenamefont
  {{Alexander}}, \citenamefont {{Biggs}}, \citenamefont {{Brandt}},
  \citenamefont {{Chapin}}, \citenamefont {{Coppin}}, \citenamefont {{Devlin}},
  \citenamefont {{Dickinson}}, \citenamefont {{Dunlop}}, \citenamefont {{Dye}},
  \citenamefont {{Eales}}, \citenamefont {{Frayer}}, \citenamefont {{Halpern}},
  \citenamefont {{Hughes}}, \citenamefont {{Ibar}}, \citenamefont
  {{Kov{\'a}cs}}, \citenamefont {{Marsden}}, \citenamefont {{Moncelsi}},
  \citenamefont {{Netterfield}}, \citenamefont {{Pascale}}, \citenamefont
  {{Patanchon}}, \citenamefont {{Rafferty}}, \citenamefont {{Rex}},
  \citenamefont {{Schinnerer}}, \citenamefont {{Scott}}, \citenamefont
  {{Semisch}}, \citenamefont {{Smail}}, \citenamefont {{Swinbank}},
  \citenamefont {{Truch}}, \citenamefont {{Tucker}}, \citenamefont {{Viero}},
  \citenamefont {{Walter}}, \citenamefont {{Wei{\ss}}}, \citenamefont
  {{Wiebe}},\ and\ \citenamefont {{Xue}}}]{IvisonEtAl2010b}%
  \BibitemOpen
  \bibfield  {author} {\bibinfo {author} {\bibfnamefont {R.~J.}\ \bibnamefont
  {{Ivison}}}, \bibinfo {author} {\bibfnamefont {D.~M.}\ \bibnamefont
  {{Alexander}}}, \bibinfo {author} {\bibfnamefont {A.~D.}\ \bibnamefont
  {{Biggs}}}, \bibinfo {author} {\bibfnamefont {W.~N.}\ \bibnamefont
  {{Brandt}}}, \bibinfo {author} {\bibfnamefont {E.~L.}\ \bibnamefont
  {{Chapin}}}, \bibinfo {author} {\bibfnamefont {K.~E.~K.}\ \bibnamefont
  {{Coppin}}}, \bibinfo {author} {\bibfnamefont {M.~J.}\ \bibnamefont
  {{Devlin}}}, \bibinfo {author} {\bibfnamefont {M.}~\bibnamefont
  {{Dickinson}}}, \bibinfo {author} {\bibfnamefont {J.}~\bibnamefont
  {{Dunlop}}}, \bibinfo {author} {\bibfnamefont {S.}~\bibnamefont {{Dye}}},
  \bibinfo {author} {\bibfnamefont {S.~A.}\ \bibnamefont {{Eales}}}, \bibinfo
  {author} {\bibfnamefont {D.~T.}\ \bibnamefont {{Frayer}}}, \bibinfo {author}
  {\bibfnamefont {M.}~\bibnamefont {{Halpern}}}, \bibinfo {author}
  {\bibfnamefont {D.~H.}\ \bibnamefont {{Hughes}}}, \bibinfo {author}
  {\bibfnamefont {E.}~\bibnamefont {{Ibar}}}, \bibinfo {author} {\bibfnamefont
  {A.}~\bibnamefont {{Kov{\'a}cs}}}, \bibinfo {author} {\bibfnamefont
  {G.}~\bibnamefont {{Marsden}}}, \bibinfo {author} {\bibfnamefont
  {L.}~\bibnamefont {{Moncelsi}}}, \bibinfo {author} {\bibfnamefont {C.~B.}\
  \bibnamefont {{Netterfield}}}, \bibinfo {author} {\bibfnamefont
  {E.}~\bibnamefont {{Pascale}}}, \bibinfo {author} {\bibfnamefont
  {G.}~\bibnamefont {{Patanchon}}}, \bibinfo {author} {\bibfnamefont {D.~A.}\
  \bibnamefont {{Rafferty}}}, \bibinfo {author} {\bibfnamefont
  {M.}~\bibnamefont {{Rex}}}, \bibinfo {author} {\bibfnamefont
  {E.}~\bibnamefont {{Schinnerer}}}, \bibinfo {author} {\bibfnamefont
  {D.}~\bibnamefont {{Scott}}}, \bibinfo {author} {\bibfnamefont
  {C.}~\bibnamefont {{Semisch}}}, \bibinfo {author} {\bibfnamefont
  {I.}~\bibnamefont {{Smail}}}, \bibinfo {author} {\bibfnamefont {A.~M.}\
  \bibnamefont {{Swinbank}}}, \bibinfo {author} {\bibfnamefont {M.~D.~P.}\
  \bibnamefont {{Truch}}}, \bibinfo {author} {\bibfnamefont {G.~S.}\
  \bibnamefont {{Tucker}}}, \bibinfo {author} {\bibfnamefont {M.~P.}\
  \bibnamefont {{Viero}}}, \bibinfo {author} {\bibfnamefont {F.}~\bibnamefont
  {{Walter}}}, \bibinfo {author} {\bibfnamefont {A.}~\bibnamefont
  {{Wei{\ss}}}}, \bibinfo {author} {\bibfnamefont {D.~V.}\ \bibnamefont
  {{Wiebe}}}, \ and\ \bibinfo {author} {\bibfnamefont {Y.~Q.}\ \bibnamefont
  {{Xue}}},\ }\href {\doibase 10.1111/j.1365-2966.2009.15918.x} {\bibfield
  {journal} {\bibinfo  {journal} {\mnras}\ }\textbf {\bibinfo {volume} {402}},\
  \bibinfo {pages} {245} (\bibinfo {year} {2010}{\natexlab{b}})}\BibitemShut
  {NoStop}%
\bibitem [{\citenamefont {{Bourne}}\ \emph {et~al.}(2011)\citenamefont
  {{Bourne}}, \citenamefont {{Dunne}}, \citenamefont {{Ivison}}, \citenamefont
  {{Maddox}}, \citenamefont {{Dickinson}},\ and\ \citenamefont
  {{Frayer}}}]{BourneEtAl2011}%
  \BibitemOpen
  \bibfield  {author} {\bibinfo {author} {\bibfnamefont {N.}~\bibnamefont
  {{Bourne}}}, \bibinfo {author} {\bibfnamefont {L.}~\bibnamefont {{Dunne}}},
  \bibinfo {author} {\bibfnamefont {R.~J.}\ \bibnamefont {{Ivison}}}, \bibinfo
  {author} {\bibfnamefont {S.~J.}\ \bibnamefont {{Maddox}}}, \bibinfo {author}
  {\bibfnamefont {M.}~\bibnamefont {{Dickinson}}}, \ and\ \bibinfo {author}
  {\bibfnamefont {D.~T.}\ \bibnamefont {{Frayer}}},\ }\href {\doibase
  10.1111/j.1365-2966.2010.17517.x} {\bibfield  {journal} {\bibinfo  {journal}
  {\mnras}\ }\textbf {\bibinfo {volume} {410}},\ \bibinfo {pages} {1155}
  (\bibinfo {year} {2011})}\BibitemShut {NoStop}%
\bibitem [{\citenamefont {{Basu}}\ \emph {et~al.}(2015)\citenamefont {{Basu}},
  \citenamefont {{Wadadekar}}, \citenamefont {{Beelen}}, \citenamefont
  {{Singh}}, \citenamefont {{Archana}}, \citenamefont {{Sirothia}},\ and\
  \citenamefont {{Ishwara-Chandra}}}]{BasuEtAl2015}%
  \BibitemOpen
  \bibfield  {author} {\bibinfo {author} {\bibfnamefont {A.}~\bibnamefont
  {{Basu}}}, \bibinfo {author} {\bibfnamefont {Y.}~\bibnamefont {{Wadadekar}}},
  \bibinfo {author} {\bibfnamefont {A.}~\bibnamefont {{Beelen}}}, \bibinfo
  {author} {\bibfnamefont {V.}~\bibnamefont {{Singh}}}, \bibinfo {author}
  {\bibfnamefont {K.~N.}\ \bibnamefont {{Archana}}}, \bibinfo {author}
  {\bibfnamefont {S.}~\bibnamefont {{Sirothia}}}, \ and\ \bibinfo {author}
  {\bibfnamefont {C.~H.}\ \bibnamefont {{Ishwara-Chandra}}},\ }\href {\doibase
  10.1088/0004-637X/803/2/51} {\bibfield  {journal} {\bibinfo  {journal}
  {\apj}\ }\textbf {\bibinfo {volume} {803}},\ \bibinfo {eid} {51} (\bibinfo
  {year} {2015})}\BibitemShut {NoStop}%
\bibitem [{\citenamefont {{Madau}}\ \emph {et~al.}(1998)\citenamefont
  {{Madau}}, \citenamefont {{Pozzetti}},\ and\ \citenamefont
  {{Dickinson}}}]{MadauEtAl1998}%
  \BibitemOpen
  \bibfield  {author} {\bibinfo {author} {\bibfnamefont {P.}~\bibnamefont
  {{Madau}}}, \bibinfo {author} {\bibfnamefont {L.}~\bibnamefont {{Pozzetti}}},
  \ and\ \bibinfo {author} {\bibfnamefont {M.}~\bibnamefont {{Dickinson}}},\
  }\href {\doibase 10.1086/305523} {\bibfield  {journal} {\bibinfo  {journal}
  {\apj}\ }\textbf {\bibinfo {volume} {498}},\ \bibinfo {pages} {106} (\bibinfo
  {year} {1998})}\BibitemShut {NoStop}%
\end{thebibliography}
\end{document}